\documentclass[a4paper,11pt]{article}

\usepackage{jheppub}

\usepackage{amsmath,amssymb,euscript}
\usepackage{slashed}
\usepackage{color}
\usepackage{accents}
\usepackage{hyperref}
\usepackage{ulem}
\usepackage{varioref}
\usepackage{xcolor}
\usepackage{xspace}
\usepackage{verbatim}
\usepackage{multirow}
\usepackage{booktabs,graphicx,tabularx,slashed}
\usepackage{mathtools}

\usepackage{tabulary}
\newcolumntype{K}[1]{>{\centering\arraybackslash}p{#1}}

\graphicspath{{figs/}}
\emergencystretch=\maxdimen
\newcommand{\tev}{\,\, \mathrm{TeV}}
\newcommand{\gev}{\,\, \mathrm{GeV}}

\newcommand{\akt}       {anti-\ensuremath{k_t}\xspace}
\newcommand{\thetaj}       {\ensuremath{\theta_{j}}\xspace}
\newcommand{\AEopt}    {\ensuremath{A_E^{\text{opt}}}\xspace}
\newcommand{\sigmaopt}    {\ensuremath{\sigma^{\text{opt}}_{\ttjet}}\xspace}
\newcommand{\ttjet}    {\ensuremath{t\bar{t}j}\xspace}
\newcommand{\ttbar}    {\ensuremath{t\bar{t}}\xspace}
\newcommand{\pt}       {\ensuremath{p_T}\xspace}
\newcommand{\ifb}       {\ensuremath{\,\mathrm{fb}^{-1}}\xspace}

\newcommand{\ETmiss}       {\ensuremath{E_T^{\text{miss}}}\xspace}
\newcommand{\Ndata}       {\ensuremath{D}\xspace}
\newcommand{\Nbkg}       {\ensuremath{B}\xspace}
\newcommand{\effReco}       {\ensuremath{\varepsilon_{\mathrm{Reco}}}\xspace}
\newcommand{\effPart}       {\ensuremath{\varepsilon_{\mathrm{Part}}}\xspace}
\newcommand{\fttbar}       {\ensuremath{f_{\ttbar}}\xspace}

\newcommand{\Secref}[1]       {Section~\ref{sec:#1}}

\newcommand{\Tabref}[1]       {Table~\ref{tab:#1}}

\begin{document}
\title{Measuring the top energy asymmetry at the LHC: QCD and SMEFT interpretations}

\author[1]{Alexander Basan,}
\author[1]{Peter Berta,}
\author[1]{Lucia Masetti,}
\author[2]{Eleni Vryonidou,}
\author[3]{Susanne Westhoff}
\affiliation[1]{PRISMA$^+$ Cluster of Excellence and Institute of Physics,  Johannes Gutenberg University Mainz, Mainz, 
 Germany}
\affiliation[2]{Theoretical Physics Department, CERN, Geneva, Switzerland}
\affiliation[3]{Institute for Theoretical Physics, Heidelberg University, Germany}

		\begin{flushright}
			P3H-20-004 \\
			CERN-TH-2020-010
		\end{flushright}
		

\abstract{The energy asymmetry in top-antitop-jet production is an observable of the top charge asymmetry designed for the LHC. We perform a realistic analysis in the boosted kinematic regime, including effects of the parton shower, hadronization and expected experimental uncertainties. Our predictions at particle level show that the energy asymmetry in the Standard Model can be measured with a significance of $3\sigma$ during Run 3, and with more than $5\sigma$ significance at the HL-LHC. Beyond the Standard Model the energy asymmetry is a sensitive probe of new physics with couplings to top quarks. In the framework of the Standard Model Effective Field Theory, we show that the sensitivity of the energy asymmetry to effective four-quark interactions is higher or comparable to other top observables and resolves blind directions in current LHC fits. We suggest to include the energy asymmetry as an important observable in global searches for new physics in the top sector.}

\maketitle
\flushbottom

\clearpage
	
	
	
\section{Introduction}\label{sec:intro}
\noindent The LHC has literally brought a quantum leap in top-quark physics. Thanks to the high precision in prediction and measurement, we can probe subtle quantum effects in top pair production and single top production, in the Standard Model (SM) and beyond~\cite{Tanabashi:2018oca}. With the large data set collected during Run 2, less frequent processes like associated top pair production with jets or with electroweak bosons have gained in importance. They allow us to examine essentially all interactions of the top quark for signs of new physics~\cite{Degrande:2010kt,Buckley:2015nca,Buckley:2015lku,Bylund:2016phk,Schulze:2016qas,Englert:2016aei,Englert:2018byk,Hartland:2019bjb,Brivio:2019ius}.

In this work we focus on top pair production in association with a hard jet. In the past this process has been investigated mostly in the Standard Model, in the context of jet radiation and the charge asymmetry in QCD~\cite{Dittmaier:2007wz,Dittmaier:2008uj,Melnikov:2010iu,Melnikov:2011qx,Alioli:2011as,Skands:2012mm,Bevilacqua:2015qha,Hoche:2016elu} or a precise determination of the top mass~\cite{Alioli:2013mxa,Fuster:2017rev,Aad:2019mkw}. We will use $t\bar t j$ production to investigate the charge asymmetry in QCD, but also to probe effective top-quark interactions with light quarks in the framework of Standard Model Effective Field Theory (SMEFT). The additional jet will prove beneficial to enhance the sensitivity to effective new top-quark interactions and to probe degrees of freedom that are difficult to access in inclusive top pair production.

A particularly sensitive probe of top interactions is the energy asymmetry in $t\bar t j$ production. The energy asymmetry is an observable of the charge asymmetry optimized for the LHC environment. It was first proposed in Ref.~\cite{Berge:2013xsa} and later computed to next-to-leading order (NLO) in QCD~\cite{Berge:2016oji}. With the data set collected at the LHC during Run 2, a measurement of the energy asymmetry with a high statistical significance is now in reach~\cite{Berge:2016oji}. The goal of our work is to perform a realistic analysis of the energy asymmetry in the regime of boosted top quarks, which is theoretically and experimentally well motivated. To estimate the impact of systematic uncertainties, we perform a full-fledged simulation at particle level that includes effects of the parton shower, hadronization and the expected selection efficiencies. This allows us to make concrete predictions for a planned measurement of the energy asymmetry in QCD with Run-2 data.

Beyond the Standard Model, $t\bar t j$ production is known to be sensitive to chiral top-quark interactions~\cite{Antunano:2007da,Ferrario:2009ee,Berge:2012rc,Chivukula:2013xla,Alte:2014toa,Brivio:2019ius}. We present the first full analysis of SMEFT contributions to $t\bar t j$ production, analyzing both charge-symmetric and charge-asymmetric observables. The main asset of $t\bar t j$ production is that the extra jet gives us access to new combinations of effective interactions that cannot be probed in inclusive $t\bar t$ production at tree level. Our goal is to assess the potential of the energy asymmetry to test these new directions in the SMEFT parameter space, and to compare it with the well-known rapidity asymmetry in inclusive top pair production.

This paper is organized as follows. Section~\ref{sec:ea-sm} is devoted to simulating a measurement of the energy asymmetry at the LHC with Run-2 data. In a first step, in Section~\ref{sec:lhc_parton} we study $t\bar t j$ production at the parton level with stable top quarks. In Section~\ref{sec:lhc_particle} we work at the particle level, which allows us to investigate event selection and reconstruction with its associated uncertainties in detail. In Section~\ref{sec:lhc-sm} we present our predictions for a measurement of the energy asymmetry with Run-2 data and make projections for Run 3 and the High-Luminosity upgrade of the LHC (HL-LHC). In Section~\ref{sec:ea-smeft} we explore the energy asymmetry as a probe of new physics in SMEFT. We discuss the chiral structure of operator contributions to $t\bar t j$ production and compare it with $t\bar t$ production in Section~\ref{sec:effective-dof}. In Section~\ref{sec:four-quark} we investigate the dependence of the $t\bar t j$ cross section and the energy asymmetry on effective top interactions with different color and chiral structures. We quantify to what precision the energy asymmetry can probe these interactions at the LHC and compare it with the rapidity asymmetry in Section~\ref{sec:bounds}, before concluding in Section~\ref{sec:conclusions}.

	\section{Energy asymmetry in the Standard Model}\label{sec:ea-sm}
	\noindent The energy asymmetry is an LHC observable of the charge asymmetry in top pair production in association with a hard jet, $pp\to t\bar t j$. To begin with, we briefly review the definition and the main features of the observable. A detailed discussion can be found in Refs.~\cite{Berge:2013xsa,Berge:2016oji}. The energy asymmetry is defined as~\footnote{Here and in what follows, we use $\sigma_S(\theta_j)$ and $\sigma_A(\theta_j)$ to refer to the differential distribution $d\sigma_{S,A}/d\theta_j$ integrated over an interval $[\thetaj^{\rm min},\thetaj^{\rm max}]$ centered around \thetaj.}
	\begin{equation}\label{eq:ea}
	A_E(\theta_j) = \frac{\sigma_{t\bar{t}j}(\theta_j,\Delta E > 0) - \sigma_{t\bar{t}j}(\theta_j,\Delta E < 0)}{\sigma_{t\bar{t}j}(\theta_j,\Delta E > 0) + \sigma_{t\bar{t}j}(\theta_j,\Delta E < 0)} =\frac{\sigma_A(\theta_j)}{\sigma_S(\theta_j)}.
	\end{equation}
    Here $\theta_j$ is the angle between the jet with the highest transverse momentum, $p_T(j_1)$, and the incoming parton $p_1$ in the partonic process $p_1 p_2\to t\bar t j$. The difference between the energies of the top and antitop quarks is defined as $\Delta E = E_t-E_{\bar t}$. Both $\theta_j$ and $\Delta E$ are defined in the $t\bar t j$ rest frame, which corresponds to the parton center-of-mass frame at leading order (LO) in QCD. The energy asymmetry is equivalent to a forward-backward asymmetry of the jet with respect to the top quark and thus probes the charge asymmetry directly at the parton level. The angular distribution $A_E(\theta_j)$ is symmetric under $\theta_j \leftrightarrow \pi - \theta_j$ and has a minimum at $\theta = \pi/2$.
	
	The energy asymmetry is mainly induced by the partonic process $qg\to t\bar{t}q$. Due to the partonic boost of the incoming quark, the jet distribution in this process is asymmetric. To reflect this feature, we define an optimized energy asymmetry as~\cite{Alte:2014toa,Berge:2016oji}
	\begin{equation}\label{eq:ea-opt}
	A_E^{\rm opt}(\theta_j) = \frac{\sigma_A(\theta_j,y_{t\bar tj}>0)+\sigma_A(\pi-\theta_j,y_{t\bar tj}<0)}{\sigma_{t\bar t j}(\theta_j,y_{t\bar tj}>0)+\sigma_{t\bar t j}(\pi-\theta_j,y_{t\bar tj}<0)} = \frac{\sigma_A^{\rm opt}(\theta_j)}{\sigma_S^{\rm opt}(\theta_j)}.
	\end{equation}
	Here $y_{t\bar{t}j}$ is the rapidity of the top-antitop-jet system, {\it i.e.}, the boost of the final state in the laboratory frame. This allows us to ``guess'' the direction of the incoming quark, which tends to be aligned with the boost of the final state. The optimized energy asymmetry has a deeper minimum than the energy asymmetry, which now lies at $\theta_j\approx 2\pi/5$. In our analysis, we will focus on this optimized energy asymmetry. Notice that the charge-symmetric cross section, $\sigma_S^{\rm{opt}}(\theta_j)$, is equivalent to the $t\bar t j$ production cross section, $\sigma_{t\bar t j}^{\rm{opt}}(\theta_j) = \sigma_S^{\rm{opt}}(\theta_j)$, and similarly $\sigma_{t\bar t j}(\theta_j) = \sigma_S(\theta_j)$. In our analysis, we will thus use $\sigma_{t\bar t j}^{\rm{opt}}$ and $\sigma_{t\bar t j}$ to denote the optimized cross section and the cross section as collider observables.
	
	In inclusive top pair production the charge asymmetry can be observed as a rapidity asymmetry~\cite{Antunano:2007da}
	\begin{align}\label{eq:ay}
	    A_{|y|} = \frac{\sigma_{t\bar t}(\Delta |y| > 0) - \sigma_{t\bar t}(\Delta |y| < 0)}{\sigma_{t\bar t}(\Delta |y| > 0) + \sigma_{t\bar t}(\Delta |y| < 0)} = \frac{\sigma_A^y}{\sigma_S^y}\,,\qquad \Delta |y| = |y_{t}| - |y_{\bar t}|\,,
	\end{align}
	where $y_t$ and $y_{\bar t}$ are the top and antitop rapidities in the laboratory frame. The currently most precise measurement of the rapidity asymmetry agrees well with the SM prediction,
	\begin{align}\label{eq:ay-exp-th}
	    A_{|y|}^{\rm exp} = (0.60\pm 0.15)\,\%~\text{\cite{ATLAS-CONF-2019-026}}\,,\qquad A_{|y|}^{\rm SM} = \left(0.66 \pm 0.06\right)\%~\text{\cite{Czakon:2017lgo}}\,.
	\end{align}
	The SM prediction has been computed at NNLO in QCD and includes electroweak contributions at NLO~\cite{Hollik:2011ps,Czakon:2017lgo}. For the energy asymmetry  electroweak contributions have not yet been investigated.
	
	In the Standard Model, $A_E$ and $A_{|y|}$ complement each other in probing the gauge structure of charge-asymmetric top pair production. While $A_{|y|}$ is induced at NLO QCD in $t\bar t$ production, $A_E$ is a LO observable in $t\bar t j$ production. Taken together, the two observables are sensitive to the interplay between real and virtual QCD effects in different kinematic regimes of top pair production. 

\subsection{LHC predictions at parton level}\label{sec:lhc_parton}
\noindent Since top quarks decay before hadronizing, the energy asymmetry (as any other top observable) needs to be reconstructed from the decay products. Before entering into the details of event selection and reconstruction, we analyze the energy asymmetry at {\it parton level}, assuming stable top quarks. This allows us to study the underlying hard process without being sensitive to effects of the parton shower or the decay of the tops.

For our numerical predictions of the cross section and the energy asymmetry at parton level, we use \texttt{MadGraph5\_aMC@NLO 2.6.5}~\cite{Alwall:2014hca} to perform fixed-order NLO QCD computations of the process $pp \rightarrow t\bar t j$ at a center-of-mass energy of $\sqrt{s}=13\tev$. Hard matrix elements have been folded with parton distribution functions (PDFs) using the interpolator \texttt{LHAPDF 6.1.6}~\cite{Buckley:2014ana}. We use the PDF set \texttt{NNPDF 3.0 NLO}~\cite{Ball:2014uwa}. Working in a factorization scheme with five active quark flavors, all quarks but the top quark are considered to be massless. The top mass is set to $m_t = 173.0\gev$, and the renormalization and factorization scales are fixed to $\mu_R = \mu_F = m_t$. The tops in the final state are kept stable, and parton shower and hadronization are not simulated. The parton-level objects excluding top quarks are clustered into jets using the \akt jet clustering algorithm~\cite{Cacciari:2008gp} with a distance parameter $R=0.4$ using~\texttt{FastJet} 3.3.1~\cite{Cacciari:2011ma} inside~\texttt{MadGraph5\_aMC@NLO}.
 
We evaluate the fixed-order predictions for the energy asymmetry $A_E^{\rm opt}$ and the $t\bar t j$ cross section in two different phase-space regions, defined by two selection criteria called \textit{loose} and \textit{boosted}. Later in our analysis we will focus on the boosted regime. Since the energy asymmetry is mostly induced by the quark-gluon initial state and increases with the energy difference $\Delta E$, it is largest in phase-space regions with boosted tops. The boost also allows us to improve the reconstruction of the top quarks and the hard jet from the final state, as we will discuss below.

In the {\it loose selection}, we only apply selection cuts on the transverse momentum (\pt) and pseudo-rapidity ($\eta$) of the jet with the highest \pt,
\begin{align}
   \text{loose:}\qquad p_T(j_1) > 100\gev\,,\quad |\eta(j_1)| < 2.5\,.
\end{align}
This jet is referred to as the \textit{associated jet}. The hard \pt cut allows us to correctly select the associated jet among the final-state products of a $t\bar t j$ event. This is crucial for the reconstruction of the energy asymmetry, as we will see in Section~\ref{sec:lhc_particle}.

For the {\it boosted selection}, we add criteria that select the phase-space region with boosted top quarks. In our analysis we focus on the single lepton channel, where one top decays leptonically ($t_\ell$) and the other one decays hadronically ($t_h$). 
 The hadronically decaying top ({\it hadronic top}) is required to be boosted with $\pt(t_h)>300\gev$ and $|\eta(t_h)| < 2.0$, so that its decay products are collimated. The leptonically decaying top ({\it leptonic top}) is required to fulfill the basic selection criteria $\pt(t_\ell)>50\gev$ and $|\eta(t_\ell)|<2.5$. In addition, the hadronic top is required to have a minimum distance $\Delta R = \sqrt{\Delta \phi^2 + \Delta \eta^2} > 1.5$ to the leptonic top and to the associated jet, respectively. In summary, we define the boosted selection as
 \begin{align}
   \text{boosted:\qquad }  p_T(j_1) & > 100\gev\,, & |\eta(j_1)| & < 2.5\,;\\\nonumber
    p_T(t_h) & > 300\gev\,, & |\eta(t_h)| & < 2.0\,,\qquad\quad \Delta R(t_h,\{t_\ell,j_1\}) > 1.5\,;\\\nonumber
    p_T(t_\ell) & > \ \,50\gev\,, & |\eta(t_\ell)| & < 2.5\,.
\end{align}
Since the top quarks are kept stable in our parton-level simulation, we randomly choose which top decays hadronically. The event rates obtained from the fixed-order computations are multiplied by the branching ratio $\mathcal{B}_{1\ell} = 0.438$ to obtain the rates in the single lepton decay channel.

\begin{figure}[!t]
	\centering
	\includegraphics[width=.47\textwidth]{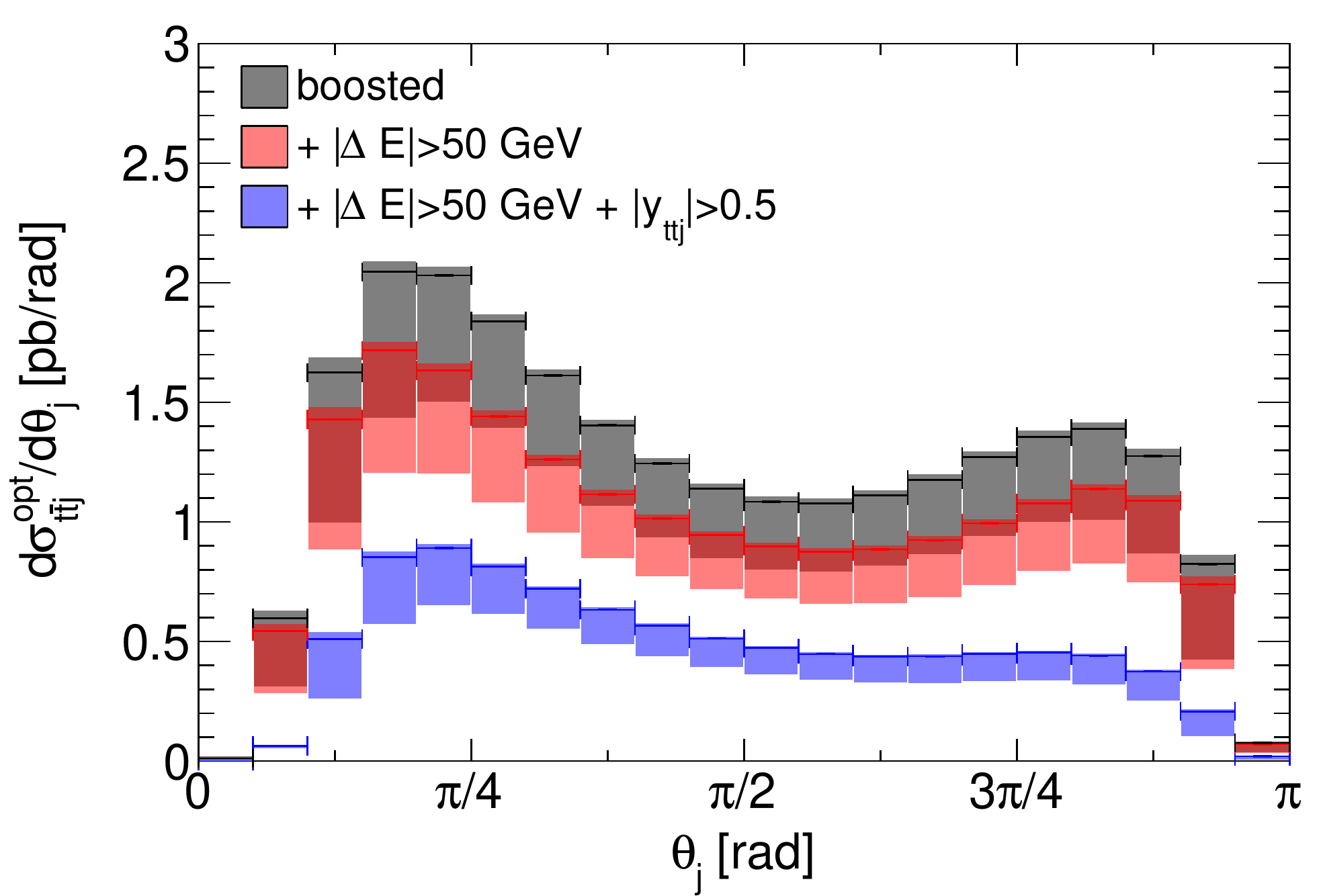}
	\hspace*{0.5cm}
	\includegraphics[width=.47\textwidth]{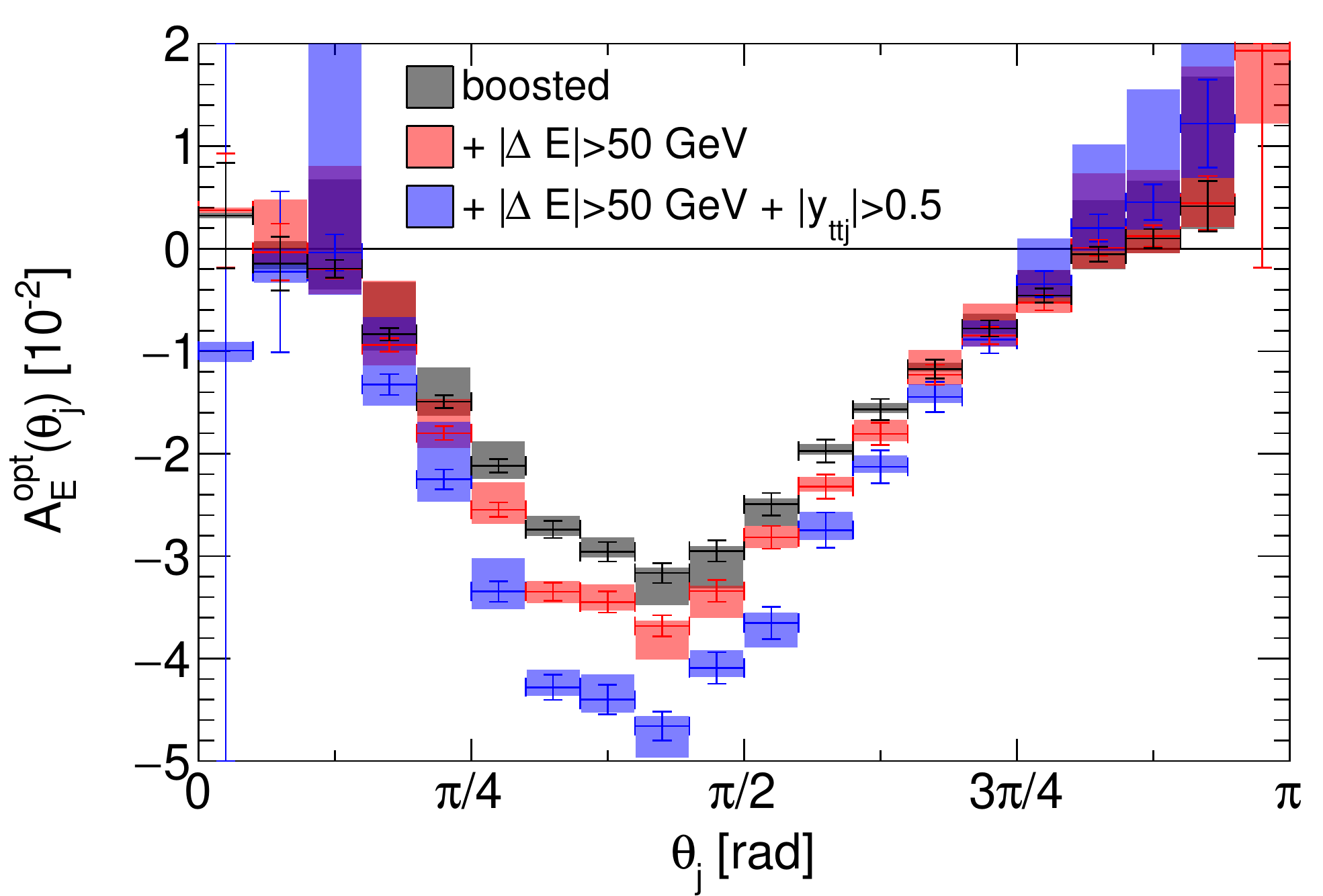}	
	\caption{Parton-level predictions at NLO QCD for the differential cross section $(d\sigma_{\ttbar j}^{\rm opt}/d\thetaj)\times \mathcal{B}_{1\ell}$ (left) and the energy asymmetry $A_E^{\rm opt}(\thetaj)$ (right) at the LHC with $\sqrt{s} = 13\tev$ as functions of the jet angle \thetaj. Shown are three different kinematic selections in the boosted top regime. Vertical error bars show statistical uncertainties from the event simulation; colored bands show the scale uncertainties.}
	\label{fig:fixedOrder}
\end{figure}

The magnitude of the energy asymmetry grows with the absolute values of the top-antitop energy difference, $|\Delta E|$, and the rapidity of the $\ttbar j$ system, $|y_{\ttbar j}|$. To focus on these phase-space regions, we set additional cuts on these two kinematic variables. In summary, we consider the following six selection regions
\begin{itemize}\label{eq:selections}
    \item loose
    \item loose \quad\ + $|\Delta E|> 50\gev$
    \item loose \quad\ + $|\Delta E|> 50\gev$ + $|y_{\ttbar j}|> 0.5$
    \item boosted
    \item boosted + $|\Delta E|> 50\gev$
    \item boosted + $|\Delta E|> 50\gev$ + $|y_{\ttbar j}|> 0.5$\,.
\end{itemize}
In Figure~\ref{fig:fixedOrder} we show the parton-level predictions at NLO QCD for the differential cross section $(d\sigma_{\ttbar j}^{\rm opt}/d\thetaj)\times \mathcal{B}_{1\ell}$ in the single lepton channel and the optimized energy asymmetry $A_E^{\rm opt}(\thetaj)$ in bins of \thetaj.
 The results are presented for the three boosted selection regions. The scale uncertainties, shown as dashed bands around the central values, are estimated from nine variations of the renormalization and factorization scales by considering all combinations from the sets $\mu_R\in \{0.5,1,2\}\,m_t$ and $\mu_F\in \{0.5,1,2\}\,m_t$.~\footnote{The upper (lower) end of the uncertainty band is defined as the maximum (minimum) obtained from these scale variations around the central value.} The energy asymmetry in the boosted selection has a minimum of \AEopt$\approx -3\%$ around $\thetaj=0.4\pi$. This minimum lies in the region of central jet emission, where theory uncertainties are well under control~\cite{Berge:2016oji}. Additional cuts on $|\Delta E|$ and $|y_{\ttbar j}|$ enhance the minimum of the asymmetry, thus potentially improving the significance of a measurement in presence of systematics-dominated uncertainties. However, this comes at the cost of reducing the cross section, leading to increased statistical uncertainties in a measurement.

In \Tabref{fixedOrder} we give numerical parton-level predictions in the six selection regions for the cross section $\sigma_{\ttbar j}$ and the energy asymmetry \AEopt in three bins of \thetaj labelled $1,2,3$ and defined by
\begin{align}\label{eq:aebins}
A_E^1:\ 0 < \theta_j < 0.3\pi\,,\qquad A_E^2: \ 0.3\pi < \theta_j < 0.7\pi\,,\qquad A_E^3:\ 0.7\pi < \theta_j < \pi.
\end{align}
The energy asymmetry increases significantly with the cuts on $\Delta E$ and $|y_{\ttbar j}|$, especially in bins 1 and 2, while the cross section decreases as discussed before. With appropriate phase-space cuts, the asymmetry in the central bin reaches $A_E^2\approx -3.6\%$, leaving a cross section of $\sigma_{\ttbar j} = 1.5\,$pb. This provides us with a good basis for a measurement in this phase-space region. In the third bin the asymmetry is small and affected by large theory uncertainties, which are mostly due to collinear jet emission. For the cross section the inclusion of NLO QCD corrections significantly reduces the scale uncertainties compared to LO QCD predictions. For the asymmetry the reduction is smaller because scale uncertainties partially cancel between the numerator and denominator of the asymmetry, but also due to a different behavior of $\sigma_A$ and $\sigma_S$ in the soft and collinear phase-space regions at NLO~\cite{Berge:2016oji}.
 
\begin{table}[!t]
\centering
	\begin{tabular}{l|ccc|c}
	\noalign{\hrule height 1pt}
	$\phantom{\Big[}$  & $A_E^1$ [$10^{-2}$] & $A_E^2$ [$10^{-2}$] & $A_E^3$ [$10^{-2}$] & $\,\sigma_{\ttbar j}$ [pb] \\
\hline
$\phantom{\Big[}$loose & $-0.48^{+0.17}_{-0.09}$  & $-1.32^{+0.08}_{-0.06}$  & $0.24^{+0.31}_{-0.14}$  & $117.5^{+6.8}_{-13.8}$ \\
$\phantom{\Big[}$loose $+|\Delta E| > 50$ & $-0.59^{+0.20}_{-0.10}$  & $-2.00^{+0.11}_{-0.08}$  & $0.29^{+0.39}_{-0.17}$  & $78.5^{+1.8}_{-3.9}$ \\
$\phantom{\Big[}$loose $+|\Delta E| > 50, |y_{t\bar{t}j}| > 0.5$ & $-1.02^{+0.18}_{-0.10}$  & $-2.52^{+0.15}_{-0.11}$  & $0.51^{+0.50}_{-0.23}$  & $41.8^{+1.0}_{-2.1}$ \\
\hline
$\phantom{\Big[}$boosted & $-1.11^{+0.38}_{-0.14}$  & $-2.42^{+0.07}_{-0.13}$  & $-0.17^{+0.46}_{-0.13}$  & $3.8^{+0.1}_{-1.1}$ \\
$\phantom{\Big[}$boosted $+|\Delta E| > 50$ & $-1.26^{+0.45}_{-0.16}$  & $-2.81^{+0.11}_{-0.07}$  & $-0.16^{+0.59}_{-0.16}$  & $3.1^{+0.1}_{-0.9}$ \\
$\phantom{\Big[}$boosted $+|\Delta E| > 50, |y_{t\bar{t}j}| > 0.5\ $ & $-1.88^{+0.60}_{-0.22}$  & $-3.58^{+0.15}_{-0.06}$  & $0.06^{+0.87}_{-0.23}$  & $1.5^{+0.0}_{-0.4}$ \\
	\noalign{\hrule height 1pt}
	\end{tabular}
	\caption{Parton-level predictions at NLO QCD for the energy asymmetry \AEopt in three \thetaj bins and for the cross section $\sigma_{\ttbar j}\times \mathcal{B}_{1\ell}$ at the LHC with $\sqrt{s} = 13\tev$. The results are shown in six selection regions; the quoted uncertainties are due to scale variations. Cuts in $|\Delta E|$ are in units of GeV.}
    \label{tab:fixedOrder}
\end{table}

\subsection{Particle-level predictions and sensitivity of an LHC measurement}\label{sec:lhc_particle}
\noindent In LHC analyses with top quarks, the results of a measurement are often not presented in terms of stable tops at parton level, but rather at {\it particle level}, where the decayed tops are reconstructed from stable final-state particles detectable by the LHC experiments. Measured rates are unfolded to the particle level and reported as observables in a fiducial phase space. Compared with a complete unfolding to parton level, the measurement uncertainties at particle level are significantly lower and ambiguities about the definition of parton-level observables are avoided.

In this section, we make predictions of the energy asymmetry at particle level to provide a sound basis for a future LHC measurement. Our goal is to assess how the energy asymmetry is modified when going from parton to particle level. In a first step, we define each final-state object at particle level and determine a fiducial phase space using particle-level objects, in a similar way as in previous \ttbar measurements at the LHC experiments. We compute  particle-level predictions of the energy asymmetry \AEopt in the fiducial phase space using NLO QCD simulations including the parton shower and hadronisation. Finally we estimate the experimental uncertainties expected in a future LHC measurement based on Run-2 data and derive projections for the expected data sets from Run 3 and the HL-LHC.

\subsubsection{Event generation}
\noindent To simulate the process $pp\rightarrow\ttjet$ at particle level, we have generated 300 million events using \texttt{Madgraph5\_aMC@NLO 2.6.5}~\cite{Alwall:2014hca} at NLO QCD interfaced with \texttt{MadSpin}~\cite{Artoisenet:2012st} for top-quark decays, including spin correlations and finite-width effects, and \texttt{Pythia 8.2.40}~\cite{Sjostrand:2014zea} for parton showering and hadronization. The entire procedure has been carried out within \texttt{MadGraph5\_aMC@NLO}, with \texttt{MC@NLO} matching. The events are analyzed with \texttt{Rivet 2.7.0}~\cite{Buckley:2010ar} using \texttt{FastJet 3.3.1}~\cite{Cacciari:2011ma} for jet clustering. Only events with one leptonic and one hadronic top are considered.

At the level of event generation we require at least one top quark with $p_T>250\gev$ and one associated jet with $p_T>70\gev$ in the hard-scattering process. These initial cuts prevent us from simulating too many events that will be rejected when applying stricter requirements on the hadronic top and on the associated jet at particle level. Our preselection is significantly looser than the requirements that define the fiducial phase space at particle level. This ensures that generation cuts do not affect our particle-level predictions.

\subsubsection{Object definition at particle level}
\noindent In our analysis, we define the objects at particle level according to the ATLAS proposal for truth particle observable definitions~\cite{ATL-PHYS-PUB-2015-013}. These definitions apply to stable particles with a mean lifetime $\tau > 30\,\text{ps}$, corresponding to a nominal decay length of $c\tau>10\,\text{mm}$. 

Electrons and muons are required to be prompt, {\it i.e.}, to be produced directly in top decays and not as secondary leptons from hadron decays. Electrons and muons from tau decays are valid prompt leptons in our analysis. Charged leptons are dressed with close-by photons, so that photons with four-momenta in a cone of $\Delta R<0.1$ around the lepton are added to the lepton four-momentum.

Two different jet definitions are used to build particle-level jets by clustering all stable particles in the event, except electrons, muons and neutrinos not coming from hadrons. The first jet definition follows the \akt algorithm with a jet cone of $R = 0.4$. These so-called \textit{small jets} are used as proxies for all partons in the final state that do not originate from the hadronic top. The second jet definition is \akt with $R = 1.0$. These \textit{large jets} are used as proxies for the boosted hadronic top. $B$ hadrons have a shorter lifetime than required by the stable-particle definition. The jet flavor is thus assigned via ghost-matching, {\it i.e.}, by including $B$ hadrons in the jet clustering algorithm with an infinitely small momentum. Any jet containing at least one $B$ hadron among its constituents is considered as a $b$-jet. Small jets are required to have $\pt>25\gev$ and $|\eta| < 2.5$. Large jets are trimmed~\cite{Krohn:2009th} as described in Ref.~\cite{ATLAS:2016vmy} and are required to have $p_T > 300\gev$ and $|\eta|< 2.0$ after trimming. Electrons and muons within a cone of $\Delta R < 0.4$ around any small jet are removed.
 
The missing transverse energy is defined as \ETmiss $= |\vec{p}_T^{\rm~miss}|$, where the transverse missing momentum $\vec{p}_T^{\rm~miss}$ is the transverse vector sum of all neutrino momenta from the hard-scattering process. In our final state with one leptonic top, the missing momentum is composed of either one neutrino from $W^+\to \ell^+\nu_\ell$ or the sum of three neutrinos from $W^+\to \tau^+\nu_\tau \to \ell^+\nu_\ell\bar{\nu}_\tau\nu_\tau$. We define the transverse momentum of the {\it particle-level neutrino} as $\vec{p}_T^{\rm~miss}$. 
 The longitudinal component of this neutrino momentum, $p_L^{\rm~miss}$, is obtained by requiring that the invariant mass of the lepton-neutrino system at particle level equals the $W$ boson mass~\cite{CERN-EP-2019-149} and that particle-level neutrino is massless. For more than one real solution of the resulting quadratic equation, we choose the result with the smaller absolute longitudinal momentum. If there is no real solution, we choose the real part of the complex solution as longitudinal neutrino momentum.

\subsubsection{Fiducial phase space}
\noindent The fiducial phase space at particle level is defined by an event selection targeting the \ttjet signal in the boosted topology. In this topology, we expect that all decay products of the hadronic top are collimated into a single large jet. Our particle-level selection corresponds closely to previous \ttbar-related LHC measurements, such as in Ref.~\cite{CERN-EP-2019-149}. In these measurements the particle-level selection criteria follow closely the selection criteria at detector level. They are devised to take account of the detector acceptance and to suppress events from background processes, while preserving as many signal events as possible.

In our selection we require exactly one lepton $\ell=e,\mu$ with $\pt>27\gev$ and $|\eta| < 2.5$. To suppress multijet background at detector level, we apply further requirements on the missing transverse momentum, $\ETmiss \ge 20 \gev$, and on the sum with the transverse mass $m_T^W$ of the $W$ boson, $\ETmiss + m_T^W \ge 60 \gev$. The $W$ momentum is defined as the four-vector sum of the lepton and neutrino momenta.

The large jet with the highest $p_T$ (referred to as $lj$) with jet mass $m\in[120,220]\gev$ is assumed to contain all the decay products of the hadronic top. It is required to be well separated from the lepton by imposing $\Delta \phi(lj,\ell) > 1.0$.

A small jet (referred to as $sj$) within a cone of $\Delta R(sj,\ell) < 2.0$ around a lepton is assumed to stem from the leptonic top. It is required to be separated from the large jet by requesting $\Delta R(sj,lj)>1.5$. If there are more than one jets fulfilling these criteria, we select the $b$-jet with the highest \pt  as our $sj$ jet. If no $b$-jet is found, we take the highest-\pt jet instead. The leptonic top is then reconstructed as the four-vector sum of the selected jet $sj$ and the $W$ boson.

The definition of the associated jet (referred to as $aj$) is devised specifically for the \ttjet process. We select the remaining small jet with the highest \pt larger than $100\gev$, required to be separated from the large jet by $\Delta R(aj,lj) > 1.5$.

\subsubsection{LHC predictions and expected experimental uncertainties}\label{sec:lhc-sm}
\begin{figure}[!t]
	\centering
	\includegraphics[width=.47\textwidth]{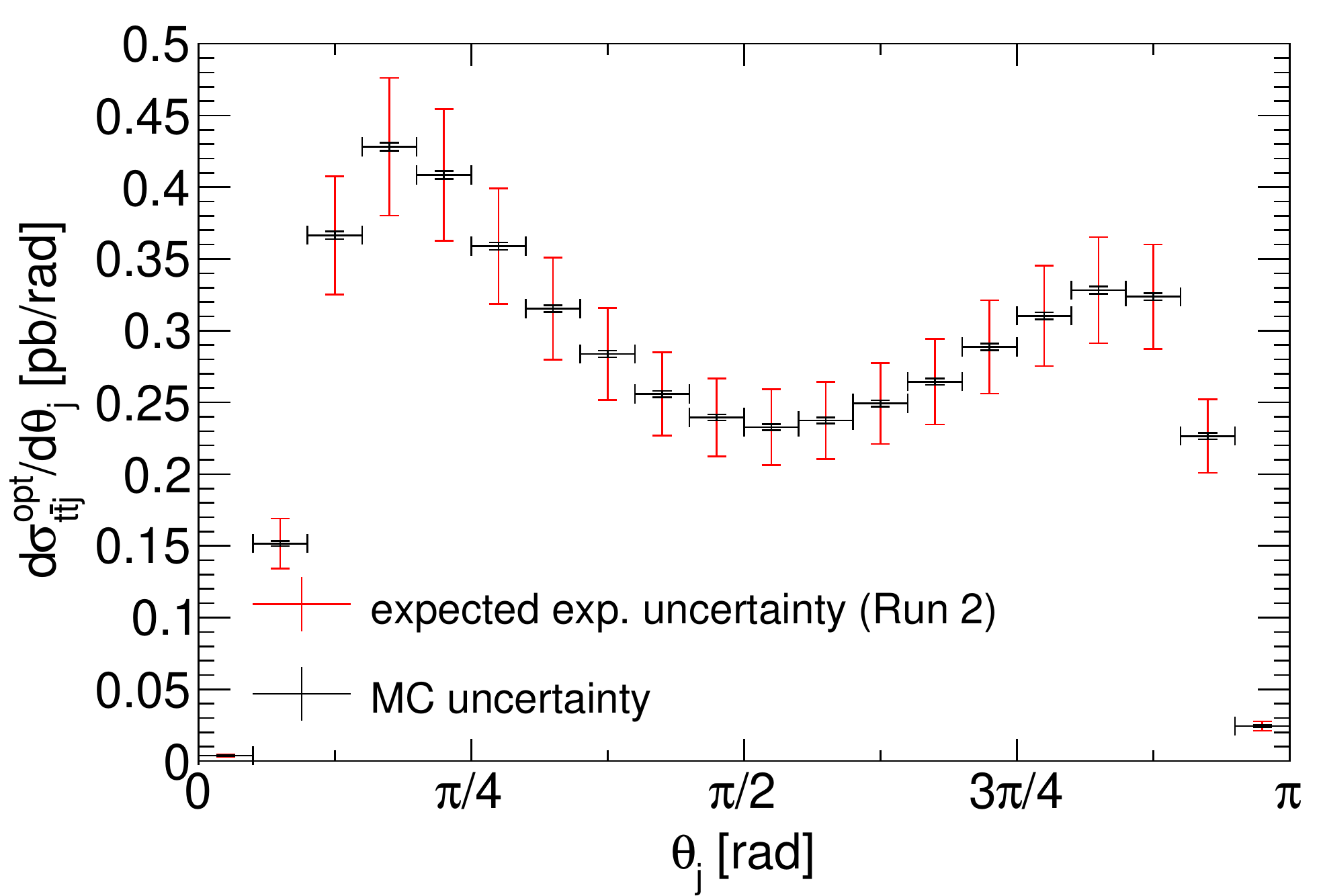}	
	\hspace*{0.5cm}
	\includegraphics[width=.47\textwidth]{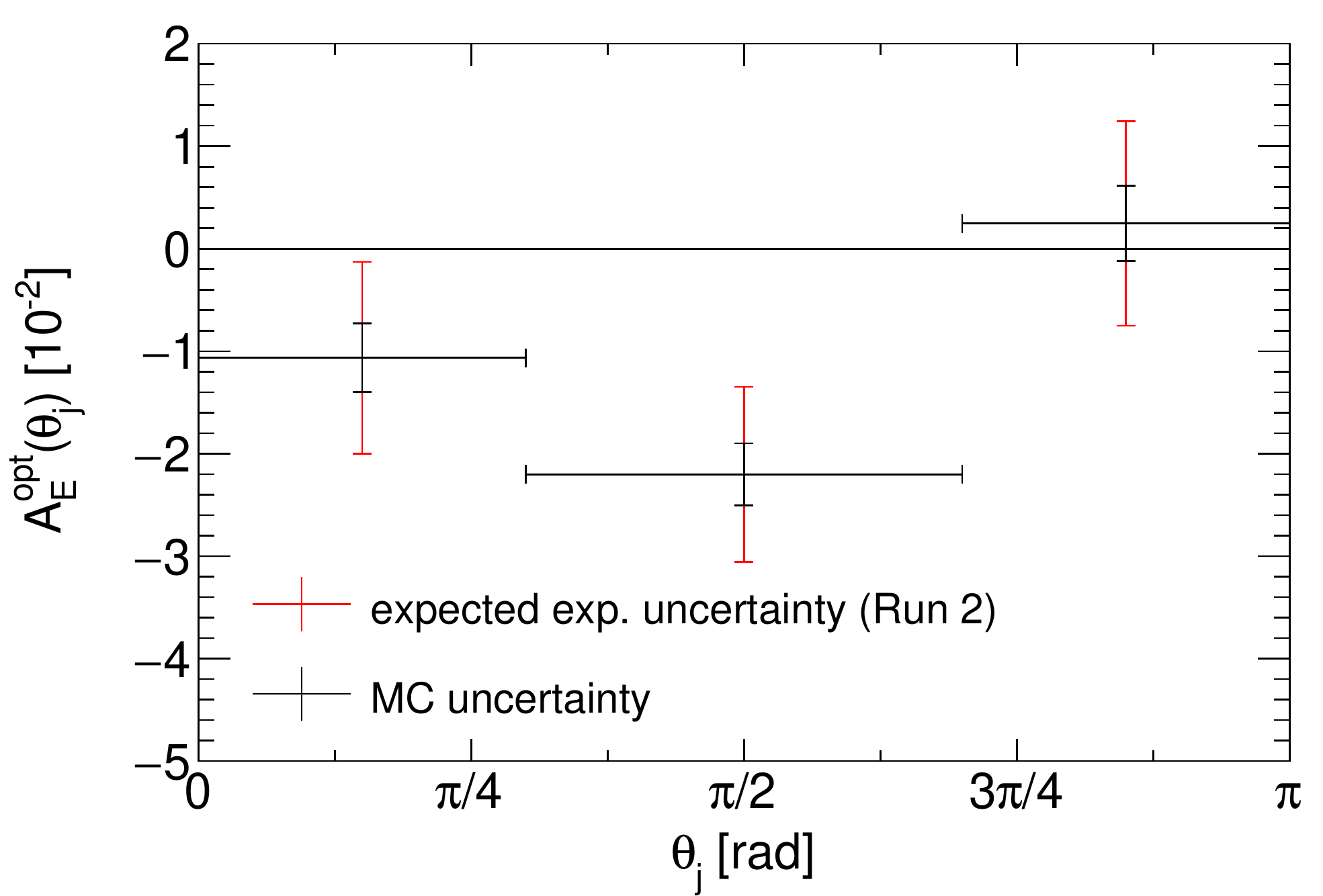}	
	\caption{Particle-level predictions of the differential cross section $d\sigma_{t\bar t j}^{\rm opt}/d\theta_j$ in the single lepton channel (left) and the energy asymmetry \AEopt (right) in three bins of \thetaj.
	The black vertical error bars show the statistical Monte-Carlo uncertainty of the prediction due to limited number of generated events. The red vertical error bars show the expected experimental uncertainties in a future LHC measurement with 139 fb$^{-1}$ of Run-2 data.}
    \label{fig:ParticleLevelExpectedMeasurement}
\end{figure}

\noindent Based on the event generation and particle-level definitions described above, we obtain predictions for the differential cross section $d\sigma_{t\bar t j}^{\rm opt}/d\theta_j$ and the energy asymmetry \AEopt in three bins of $\thetaj$ at particle level in the fiducial region. Our results are shown in Figure~\ref{fig:ParticleLevelExpectedMeasurement}. The statistical uncertainty due to the limited number of simulated events, dubbed {\it Monte-Carlo (MC) uncertainty}, is shown as black vertical lines.
 We find that the shape of the angular distributions $d\sigma_{t\bar t j}^{\rm opt}/d\theta_j$ and $A_E^{\rm opt}(\theta_j)$ remains close to the parton-level predictions from~\Secref{lhc_parton}, which targets the same boosted phase-space region as our particle-level selection. This shows that effects of the parton shower, hadronization and event reconstruction do not affect the relevant kinematics for these observables. The total fiducial cross section in our particle-level selection is roughly four times smaller than in the parton-level boosted selection. This is due to the tighter selection criteria for the particle-level phase space, which are expected to be applied in a real detector environment. The predicted energy asymmetry in the central \thetaj bin, $A_E^2 = -2.2\cdot 10^{-2}$, has roughly the same magnitude as at parton level, $A_E^2 = -2.4\cdot 10^{-2}$ (see Table~\ref{tab:fixedOrder}). This is a positive sign indicating that the energy asymmetry can be measured with a comparable magnitude in a real-detector environment.

 To assess the sensitivity of an LHC measurment to the energy asymmetry, in Figure~\ref{fig:ParticleLevelExpectedMeasurement} we also show the {\it expected experimental uncertainty} (red) corresponding to a Run-2 data set from an integrated luminosity of $L=139\ifb$.
 Details on our estimation of this uncertainty are given in Appendix~\ref{app:uncertainties}. We identify two main sources of experimental uncertainties: the statistical uncertainty $\Delta A_E^{\rm stat}$ due to the detected number of events; and the systematic uncertainty $\Delta A_E^{\rm bkg}$ on the estimated number of background events. Using realistic estimates of these uncertainties from the appendix, we obtain the resulting uncertainties on the energy asymmetry from error propagation as
\begin{equation}\label{eq:exp-unc}
\Delta A_E^{\text{stat}}(\thetaj)\approx\frac{1.4}{ \sqrt{L\,\sigma_{t\bar t j}^{\rm opt}(\thetaj)}}\,,\qquad\quad 
\Delta A_E^{\text{bkg}}(\thetaj)\approx 0.018\cdot|A_{E,\mathrm{meas}}^{\text{opt}}(\thetaj)|\,.
\end{equation}
In Table~\ref{tab:ParticleLevelExpectedMeasurement} we give numerical predictions for the optimized energy asymmetry in three \thetaj bins and the $t\bar t j$ cross section in the fiducial region at particle level, together with the expected experimental uncertainties. The uncertainties are obtained for three different values of integrated luminosity corresponding to the available data set from LHC Run 2, the expected combined data sets from LHC Run 2 and Run 3, and the data set expected from the HL-LHC. For our SMEFT analysis in \Secref{bounds}, we will use the expected experimental uncertainties for the Run-2 data set.

\begin{table}[!t]
\centering
	\begin{tabular}{l|ccc|c}
		\noalign{\hrule height 1pt}
\ luminosity $L\,[\ifb]$$\phantom{\Big[}$  & $A_E^1\ [10^{-2}]$ & $A_E^2\ [10^{-2}]$ & $A_E^3\ [10^{-2}]$ & $\sigma_{t\bar{t}j}$ [pb]\\
\hline
\ 139 (LHC Run 2)$\phantom{\Big[}$  & $-1.06\pm0.93$ & $-2.20\pm0.85$ & $0.25\pm1.00\ $ & $\ 0.83 \pm 0.04$ \\
\ 300 (LHC Run 2+3)$\phantom{\Big[}$  & $-1.06\pm0.64$ & $-2.20\pm0.59$ & $0.25\pm0.68\ $ & $\ 0.83 \pm 0.04$ \\
\ 3000 (HL-LHC)$\phantom{\Big[}$  & $-1.06\pm0.21$ & $-2.20\pm0.21$ & $0.25\pm0.22\ $  & $\ 0.83 \pm 0.04$\\
	\noalign{\hrule height 1pt}
	\end{tabular}
	\caption{Particle-level predictions for the energy asymmetry \AEopt in three \thetaj bins and for the cross section $\sigma_{t\bar t j}$ in the fiducial region at the LHC with $\sqrt{s}=13\tev$. The quoted errors are the expected experimental uncertainties in a measurement using data sets of integrated luminosity $L$.  \label{tab:ParticleLevelExpectedMeasurement}}
\end{table}

For a given luminosity the expected absolute statistical uncertainty $\Delta A_E^{\text{stat}}(\thetaj)$ from Eq.~\eqref{eq:exp-unc} is roughly the same for the three asymmetry bins $A_E^1$, $A_E^2$ and $A_E^3$. The reason is the similar cross section $\sigma_{t\bar t j}^{\rm opt}(\theta_j)$ in all three \thetaj bins. Since the absolute statistical uncertainty scales as $1/\sqrt{L}$, it can be significantly lowered at the HL-LHC with a data set that is roughly twenty times larger than what is currently available from Run 2.

In all three luminosity scenarios and \thetaj bins, the expected experimental uncertainty on the asymmetry is statistics-dominated. The largest absolute background uncertainty is observed for the central bin 2, where $\Delta A_{E}^{\text{bkg}}(\thetaj) \approx 0.04\%$. This is due to the fact that the background uncertainty scales linearly with the asymmetry itself, see Eq.~\eqref{eq:exp-unc}. Compared to the statistical uncertainty, however, the background uncertainty on the energy asymmetry is subdominant even in the Run-2 scenario. The smallness of the background uncertainty is partly due to our assumption of symmetric background with respect to $\Delta E$, which ensures a cancellation of background-related uncertainties in the numerator of the asymmetry, see  Eq.~\eqref{eq:ea-opt-rewritten}. In reality additional systematic uncertainties could arise from background processes like $W$+jets, which feature an intrinsic energy asymmetry. To estimate such uncertainties, a dedicated simulation of these background processes would be required, but is beyond the scope of our analysis.

For the cross section, we expect the experimental uncertainty to be systematics-dominated already with Run-2 data. The total experimental uncertainty therefore does not change significantly with a larger data set.

In summary, our predictions show good prospects to measure the energy asymmetry at the LHC. With data from Runs 2 and 3, we expect a significance of more than 3 standard deviations from zero. At the HL-LHC statistical limitations are overcome and the significance is increased to more than $5\sigma$. While additional sources of systematic uncertainties would reduce the sensitivity, there is much room to optimize the event selection and reconstruction for the \ttjet final state. We therefore believe that the quoted significance is a realistic goal for future measurements.

\section{Energy asymmetry in the Standard Model Effective Field Theory}\label{sec:ea-smeft}
\noindent Heavy new particles with top-quark couplings can drastically change the energy asymmetry compared to its SM prediction. Our analysis of the energy asymmetry in the Standard Model predicts that it can be measured at the LHC with less than $50\,\%$ uncertainty already with the present data set from Run 2, see Table~\ref{tab:ParticleLevelExpectedMeasurement}. Given that $t\bar t j$ observables have a high theoretical sensitivity to new top interactions~\cite{Ferrario:2009ee,Berge:2012rc,Alte:2014toa}, even a loose measurement of the energy asymmetry will provide us with interesting information about new physics at high scales. To provide a model-independent theoretical framework for new-physics interpretations of a future measurement, we perform a detailed analysis of the energy asymmetry in SMEFT.

At the LHC, generic effects of new physics at high scales $\Lambda \gtrsim 1\,\text{TeV}$ can be described by an effective Lagrangian
	\begin{align}
\mathcal{L}_\text{eff} = \sum_k \left(\frac{C_k}{\Lambda^2}\,^\ddagger O_k + \text{h.c.} \right) 
               + \sum_l \frac{C_l}{\Lambda^2}\,O_l\,,
\end{align}
	with Wilson coefficients $C_k$ of local operators $O_k$. Non-hermitian operators are denoted as $^\ddagger O$. The sum runs over all SM gauge invariant operators at mass dimension 6. We neglect operators of higher mass dimension in the SMEFT expansion and assume CP conservation, which implies that all Wilson coefficients are real. Furthermore, we request a $U(2)_q\times U(2)_u\times U(2)_d$ flavor symmetry among quarks of the first and second generation. Left- and right-handed quarks of the first two generations and the third generation are denoted by
	\begin{align}
	q_i & = (u^i_L,d^i_L),\qquad \,u_i = u^i_R,\,d_i=d^i_R,\qquad i=1,2\\\nonumber
	Q & = (t_L,b_L),\qquad \quad t = t_R,\,b=b_R\,.
	\end{align}
	Under these assumptions, there are 15 independent operators with top quarks that contribute to $t\bar t$ and $t\bar t j$ production at tree level~\cite{AguilarSaavedra:2018nen}:
	\begin{itemize}
		\item 8 four-quark operators with Lorentz structures $LL$ and $RR$,
		\begin{align}\label{eq:llrr}
		O_{Qq}^{1,8} & = (\bar{Q}\gamma_\mu T^A Q)(\bar{q}_i\gamma^\mu T^A q_i), & O_{Qq}^{1,1} & = (\bar{Q}\gamma_\mu Q)(\bar{q}_i\gamma^\mu q_i),\\\nonumber
		O_{Qq}^{3,8} & = (\bar{Q}\gamma_\mu T^A\tau^I Q)(\bar{q}_i\gamma^\mu T^A \tau^I q_i), & O_{Qq}^{3,1} & = (\bar{Q}\gamma_\mu\tau^I Q)(\bar{q}_i\gamma^\mu\tau^I q_i),\\\nonumber
		O_{tu}^8 & = (\bar{t}\gamma_\mu T^A t)(\bar{u}_i\gamma^\mu T^A u_i), & O_{tu}^1 & = (\bar{t}\gamma_\mu t)(\bar{u}_i\gamma^\mu u_i),\\\nonumber
		O_{td}^8 & = (\bar{t}\gamma^\mu T^A t)(\bar{d}_i\gamma_\mu T^A d_i), & O_{td}^1 & = (\bar{t}\gamma^\mu t)(\bar{d}_i\gamma_\mu d_i),
		\end{align}
		\item 6 four-quark operators with Lorentz structures $LR$ and $RL$,
		\begin{align}\label{eq:lrrl}
		O_{Qu}^8 & = (\bar{Q}\gamma^\mu T^A Q)(\bar{u}_i\gamma_\mu T^A u_i), & O_{Qu}^1 & = (\bar{Q}\gamma^\mu Q)(\bar{u}_i\gamma_\mu u_i),\\\nonumber
		O_{Qd}^8 & = (\bar{Q}\gamma^\mu T^A Q)(\bar{d}_i\gamma_\mu T^A d_i), & O_{Qd}^1 & = (\bar{Q}\gamma^\mu Q)(\bar{d}_i\gamma_\mu d_i),\\\nonumber
		O_{tq}^8 & = (\bar{q}_i\gamma^\mu T^A q_i)(\bar{t}\gamma_\mu T^A t), & O_{tq}^1 & = (\bar{q}_i\gamma^\mu q_i)(\bar{t}\gamma_\mu t).
		\end{align}
		\item 1 tensor operator that modifies the top-gluon interaction,
		\begin{align}
		^\ddagger O_{tG} = (\bar{Q}\sigma^{\mu\nu} T^A t)\,\widetilde{\phi}\,G_{\mu\nu}^A.
		\end{align}
	\end{itemize}
	The degrees of freedom that contribute to top pair production are conveniently expressed in terms of vector and axial-vector currents. For color-octet operators with up quarks, we define the combinations~\cite{Brivio:2019ius}
\begin{align}
4\,C_{VV}^{u,8} & =  C_{Qq}^{1,8} + C_{Qq}^{3,8} + C_{tu}^8 + C_{tq}^8 + C_{Qu}^8 \notag \\
4\,C_{AA}^{u,8} & = C_{Qq}^{1,8} + C_{Qq}^{3,8} + C_{tu}^8 - C_{tq}^8 - C_{Qu}^8 \notag \\
4\,C_{AV}^{u,8} & = - \big(C_{Qq}^{1,8} + C_{Qq}^{3,8}\big) + C_{tu}^8 + C_{tq}^8 - C_{Qu}^8 \notag \\
4\,C_{VA}^{u,8} & = - \big(C_{Qq}^{1,8} + C_{Qq}^{3,8}\big) + C_{tu}^8 - C_{tq}^8 + C_{Qu}^8 \; .
\label{eq:vv-aa}
\end{align}
Analogous definitions hold for color-singlet operators, changing the index $8\to 1$. Combinations for operators with down quarks are obtained by replacing the index $u \to d$ and
$+ C_{Qq}^{3,8} \to - C_{Qq}^{3,8}$ in Eq.~\eqref{eq:vv-aa}. In LHC observables the relative contributions of operators with up or down quarks depend on the parton distributions inside the proton. For $q\bar q-$ and $qg-$initiated processes the relevant ratios are
\begin{align}
r_{q\bar q}(x) = \frac{f_u(x)f_{\bar u}(s/(xS))}{f_d(x)f_{\bar d}(s/(xS))}\,, \qquad\quad r_{qg}(x) = \frac{f_u(x)f_{g}(s/(xS))}{f_d(x)f_{g}(s/(xS))}\,,
\end{align}
where $f_p(x,s)$ denotes the distribution of parton $p$ with momentum fraction $x$ inside the proton, and $\sqrt{s}$ and $\sqrt{S}$ are the partonic and hadronic center-of-mass energy, respectively. Observables in $\ttbar j$ production  schematically probe the combinations ($r = r_{q\bar q}, r_{qg}$)
\begin{align}
r\, C_{VV}^{u,8} + C_{VV}^{d,8} = (r+1)(C_{Qq}^{1,8} + C_{tq}^8) + (r-1) C_{Qq}^{3,8} + r (C_{tu}^8 + C_{Qu}^8) + (C_{td}^8 + C_{Qd}^8)\,,
\end{align} where $r$ denotes the integrated form of the parton distribution ratios. In inclusive observables the ratio $r\approx 2$ reflects the number of valence quarks inside the proton. In differential distributions and asymmetries the value of $r$ depends on the relevant phase-space region, so that the relative contribution of up- and down-quark operators can vary. Similar combinations apply for color-singlets and for $C_{AA},C_{AV},C_{VA}$.

\subsection{Effective degrees of freedom in top-antitop-jet production}\label{sec:effective-dof}
\noindent To set the basis for our numerical analysis, we determine which combinations of operators can be probed as degrees of freedom in $t\bar t j$ observables. To this end we distinguish between charge-symmetric and charge-asymmetric observables,
\begin{align}
d\sigma_S & = d\sigma\big(t(p_1),\bar{t}(p_2)\big) + d\sigma\big(t(p_2),\bar{t}(p_1)\big)\,, \notag \\
d\sigma_A & = d\sigma\big(t(p_1),\bar{t}(p_2)\big) - d\sigma\big(t(p_2),\bar{t}(p_1)\big)\,,
\label{eq:sym-asym}
\end{align}
where $p_1$ and $p_2$ are the four-momenta of the top quarks in a certain phase-space region. Observables like the total cross section or the top-antitop invariant mass distribution are symmetric under top-antitop interchange, while observables of the charge asymmetry are antisymmetric. In terms of Wilson coefficients, these observables can be written as
\begin{align}\label{eq:obs-smeft}
    \sigma & = \sigma_S^{\rm SM} + \sum_k\,C_k\sigma_S^k + \sum_{k\le l} C_k C_l\,\sigma_S^{kl}\,,\\\nonumber
    A & = \frac{\sigma_A}{\sigma_S} = 
\frac{ \sigma_A^{\text{SM}} + \sum_k C_k  \sigma_A^{k} 
	+ \sum_{k\le l} C_k C_l \, \sigma_A^{kl}}{ \sigma_S^{\text{SM}} + \sum_k C_k \sigma_S^{k} +  \sum_{k\le l} C_k C_l\, \sigma_S^{kl}}\,,
\end{align}
where we have defined the Wilson coefficients $C_k,C_l$ in units of $\Lambda^{-2} = 1\,\text{TeV}^{-2}$, and $\sigma_S^{\rm SM}$ and $\sigma_A^{\rm SM}$ are the charge-symmetric and -asymmetric cross sections in the Standard Model. The cross sections $\sigma_{S,A}^k$ and $\sigma_{S,A}^{kl}$ correspond to operator interference with QCD at $\mathcal{O}(\Lambda^{-2})$ and operator interference at $\mathcal{O}(\Lambda^{-4})$, respectively. Depending on the observable, $\sigma_S$ and $\sigma_A$ correspond to (see Eqs.~\eqref{eq:ea},\,\eqref{eq:ea-opt},\,\eqref{eq:ay})
\begin{align}
A_E(\theta_j): & \qquad \sigma_{S,A}(\theta_j) = \sigma_{t\bar t j}(\theta_j, \Delta E > 0) \pm \sigma_{t\bar t j}(\theta_j, \Delta E < 0)\,,\\\nonumber 
A_E^{\rm opt}(\theta_j): & \qquad \sigma_{S,A}^{\rm opt}(\theta_j) = \sigma_{S,A}(\theta_j,y_{t\bar{t}j} > 0) + \sigma_{S,A}(\pi - \theta_j,y_{t\bar{t}j} < 0)\,,\\\nonumber
A_{|y|}: & \qquad \sigma_{S,A}^y \quad \ \ = \sigma_{t\bar t}(\Delta |y| > 0) \pm \sigma_{t\bar t}(\Delta |y| < 0)\,.
\end{align}
In Table~\ref{tab:tt-ttj-ops} we show the combinations of four-quark operator coefficients that contribute to top-antitop production (left) and top-antitop-jet production (right).
\begin{table}[t]
\centering
\renewcommand{\arraystretch}{1.3}
\begin{tabular}{c|l|l}
\noalign{\hrule height 1pt}
& $\phantom{\Big]} t\bar t$ & $\ t\bar t j$ \\
\noalign{\hrule height 1pt}
$\, \sigma_S^k \, $ & $\phantom{\Big]} C_{VV}^{q,8}$ & $\ C_{VV}^{q,8},\quad C_{AA}^{q,8} + \frac{4}{3}\,C_{AA}^{q,1}$ \\
\hline
\multirow{3}{*}{$\sigma_S^{kl}$}
  & $\phantom{\Big]} |C_{V+A}^{q,8}|^2 + \frac{9}{2}\,|C_{V+A}^{q,1}|^2$ & $\ |C_{VV}^{q,8}|^2 + |C_{VA}^{q,8}|^2,\quad |C_{AA}^{q,8}|^2 + |C_{AV}^{q,8}|^2$ \\
  & $\phantom{\Big]} |C_{V-A}^{q,8}|^2 + \frac{9}{2}\,|C_{V-A}^{q,1}|^2$ & $|C_{VV}^{q,8}|^2 + |C_{VA}^{q,8}|^2 + \frac{3}{2}\big(|C_{VV}^{q,1}|^2 + |C_{VA}^{q,1}|^2\big)$ \\
  & $\phantom{\Big]}$ & $|C_{AA}^{q,8}|^2 + |C_{AV}^{q,8}|^2 + \frac{3}{2}\big(|C_{AA}^{q,1}|^2 + |C_{AV}^{q,1}|^2\big)$ \\
  & $\phantom{\Big]}$ & $\ 2C_{VV}^{q,8} C_{AA}^{q,8} + \frac{4}{3}\big(C_{VV}^{q,1} C_{AA}^{q,8} + C_{VV}^{q,8} C_{AA}^{q,1}\big)$ \\
\noalign{\hrule height 1pt}
$\sigma_A^k$ & $\phantom{\Big]} C_{AA}^{q,8}$ & $\ C_{AA}^{q,8},\quad C_{VV}^{q,8} + \frac{4}{3}\,C_{VV}^{q,1}$ \\
\hline
\multirow{3}{*}{$\sigma_A^{kl}$} & $\phantom{\Big]} C_{VV}^{q,8} C_{AA}^{q,8} + C_{VA}^{q,8} C_{AV}^{q,8}$ & $\ C_{VV}^{q,8} C_{AA}^{q,8} + C_{VA}^{q,8}C_{AV}^{q,8}$\\
  & $\phantom{\Big]}\, + \frac{9}{2}\big(C_{VV}^{q,1} C_{AA}^{q,1} + C_{VA}^{q,1} C_{AV}^{q,1}\big)$ & $\ C_{VV}^{q,8} C_{AA}^{q,8} + C_{VA}^{q,8}C_{AV}^{q,8} + \frac{3}{2}\big(C_{VV}^{q,1} C_{AA}^{q,1} + C_{VA}^{q,1}C_{AV}^{q,1}\big)$ \\
  & $\phantom{\Big]}$ & $\ |C_{VV}^{q,8}|^2 + |C_{VA}^{q,8}|^2 + \frac{4}{3}\,2\big(C_{VV}^{q,1}C_{VV}^{q,8} + C_{VA}^{q,1}C_{VA}^{q,8}\big)$ \\
  & $\phantom{\Big]}$ & $\ |C_{AA}^{q,8}|^2 + |C_{AV}^{q,8}|^2 + \frac{4}{3}\,2\big(C_{AA}^{q,1}C_{AA}^{q,8} + C_{AV}^{q,1}C_{AV}^{q,8}\big)$ \\
\noalign{\hrule height 1pt}
\end{tabular}
\caption{\label{tab:tt-ttj-ops}Four-quark operator contributions to $t\bar t$ production (left) and $t\bar t j$ production (right) at tree level at $\mathcal{O}(\Lambda^{-2})$ and $\mathcal{O}(\Lambda^{-4})$, denoted as $\sigma_{S,A}^k$ and $\sigma_{S,A}^{kl}$, respectively.}
\end{table}
 It is apparent that $t\bar t$ production is sensitive to five combinations of Wilson coefficients~\cite{Brivio:2019ius}, including
	\begin{align}
|C_{V+A}^{q,\alpha}|^2 & = |C_{VV}^{q,\alpha}|^2 + |C_{VA}^{q,\alpha}|^2 + |C_{AA}^{q,\alpha}|^2 + |C_{AV}^{q,\alpha}|^2\,, \phantom{\Big]} \notag \\
& \stackrel{q=u}{=} \left(|C_{Qq}^{1,\alpha} + C_{Qq}^{3,\alpha}|^2 + |C_{tu}^\alpha|^2 + |C_{tq}^\alpha|^2 + |C_{Qu}^\alpha|^2\right)/4 \\
|C_{V-A}^{q,\alpha}|^2 & = |C_{VV}^{q,\alpha}|^2 + |C_{VA}^{q,\alpha}|^2 - |C_{AA}^{q,\alpha}|^2 - |C_{AV}^{q,\alpha}|^2 \notag \\
& \stackrel{q=u}{=} \left( (C_{Qq}^{1,\alpha} + C_{Qq}^{3,\alpha}) C_{tq}^\alpha  + C_{tu}^\alpha C_{Qu}^\alpha \right) / 2 \,,
\label{eq:chiral_coeff}
\end{align}
where $\alpha = 1,8$ and $q=u,d$ in $C^{q,\alpha}$. For the sake of clarity, we do not show contributions with odd numbers of $V$ or $A$ currents like $C_{VV}C_{VA}$ or $C_{AA}C_{AV}$, but include them in our numerical analysis.

In contrast, $t\bar{t} j$ production has a much richer structure in SMEFT than $t\bar t$ production. Part of the structure has been analyzed for heavy color-octet bosons in Refs.~\cite{Ferrario:2009ee,Berge:2012rc}. Here we discuss the degrees of freedom for the energy asymmetry in detail. In Figure~\ref{fig:ttbj-diagrams} we show examples of diagrams that generate the energy asymmetry through initial-state radiation (ISR-ISR) (left) or through the interference of initial- and final-state radiation (ISR-FSR) (right). Charge-asymmetric contributions are generated either from an asymmetric Lorentz structure or an asymmetric color structure of the corresponding matrix elements~\cite{Berge:2012rc}.
\begin{figure}[!t]
	\centering
	\includegraphics[width=.42\textwidth]{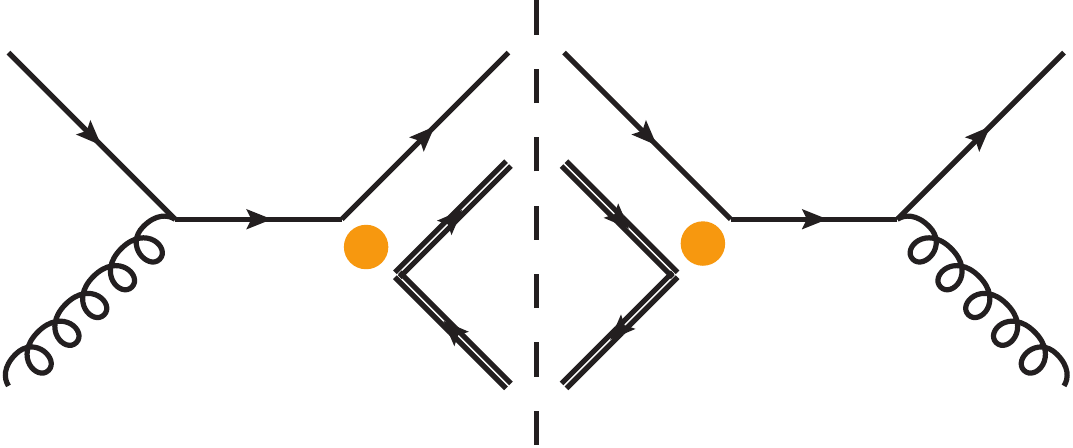}
	\hspace*{1.5cm}
	\includegraphics[width=.36\textwidth]{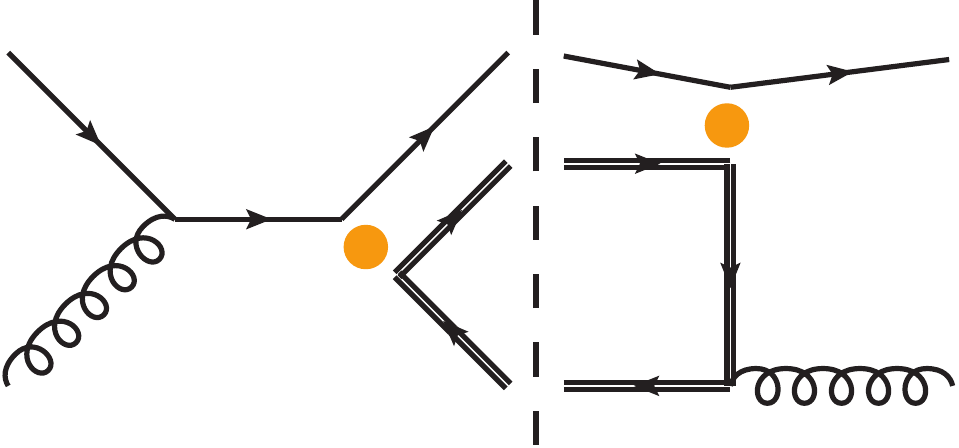}	
	\caption{Operator contributions to the charge asymmetry in $t\bar t j$ production from initial-state radiation (left) and interference of initial- and final-state radiation (right). Single lines represent light quarks, double lines represent top quarks. Orange dots indicate the insertion of a four-quark operator or of a gluon. The dashed line symbolizes the interference $\mathcal{M}_1\mathcal{M}_2^\ast$ of the two $qg \to t\bar t q$ amplitudes $\mathcal{M}_1$ and $\mathcal{M}_2^\ast$ to its left and right sides.}
	\label{fig:ttbj-diagrams}
\end{figure}

In QCD, ISR-ISR and FSR-FSR interference are symmetric under top-antitop interchange both in Lorentz and in color structure, so that they do not induce a charge asymmetry. ISR-FSR interference, in turn, has an asymmetric Lorentz structure. The color structure splits into a symmetric part $d_{abc}^2$ and an asymmetric part $f_{abc}^2$, where
   \begin{align}
       d_{abc} = 2{\rm Tr}[\{T^a,T^b\},T^c],\qquad f_{abc} = -2i{\rm Tr}[[T^a,T^b],T^c]\,, 
   \end{align}
 and $T^i$ are the eight $SU(3)_c$ generators. The QCD contribution to the charge asymmetry is thus proportional to $d_{abc}^2$~\cite{Kuhn:1998kw,Kuhn:2011ri}.

In SMEFT, operator insertions with axial-vector currents can change the Lorentz structure of the matrix elements and thus induce additional contributions to the charge asymmetry. ISR-ISR or FSR-FSR interference can have an asymmetric Lorentz structure from operator insertions with one axial-vector current on each quark line. For instance, $C_{AA}-{\rm QCD}$ interference or $C_{AA}-C_{VV}$ interference induce a contribution to the charge asymmetry. In ISR-FSR interference, the charge asymmetry can be generated either from vector operators that preserve the asymmetric Lorentz structure, or from axial-vector operators that combine a symmetric Lorentz structure with an asymmetric color structure $f_{abc}^2$. Examples are $C_{VV}-C_{VV}$ interference $\sim d_{abc}^2$ or $C_{VV}-C_{AA}$ interference $\sim f_{abc}^2$. These examples illustrate the variety of contributions to the energy asymmetry shown in Table~\ref{tab:tt-ttj-ops}.

\subsection{Properties of four-quark operators}\label{sec:four-quark}
\noindent The large number of effective degrees of freedom suggests that $t\bar t j$ observables have a good potential to probe four-quark operators with different chiral and color structures. Here we will demonstrate this potential for the cross section and the energy asymmetry in $t\bar{t}j$ production.

To illustrate the dependence of $\sigma_{t\bar tj}$ and $A_E^{\rm opt}$ on the different Wilson coefficients, we consider two pairs of four-quark operators
\begin{align}\label{eq:op-set}
    \{O_{Qq}^{1,1},\ O_{tq}^1\} \sim \{LL,RL\}\qquad \text{and} \qquad \{O_{Qq}^{1,1},\ O_{tu}^1\} \sim \{LL,RR\}\,.
\end{align}
With these two sets we can study the impact of the top chirality on the observables. Changing $LL \to LR$ would give the same results, because we cannot distinguish the chirality of light quarks due to their small mass. With the operators in Eq.~\eqref{eq:op-set}, the following relations apply
\begin{align}\label{eq:relations}
\{O_{Qq}^{1,1},\ O_{tq}^1\}: & \quad C_{VA} = - C_{VV},\ C_{AV} = - C_{AA}\,,\\\nonumber
\{O_{Qq}^{1,1},\ O_{tu}^1\}: & \quad C_{AA} = C_{VV},\ C_{AV} = C_{VA}\,,
\end{align}
which reduces the degrees of freedom in Table~\ref{tab:tt-ttj-ops}. For $\{O_{Qq}^{1,1},\ O_{tq}^1\}$, the $t\bar t j$ cross section and the energy asymmetry depend on the Wilson coefficients as
\begin{align}\label{eq:obs-in-smeft}
    \sigma_{t\bar t j}^{(\text{opt})} & = \sigma_S^{\rm SM} + (C_{Qq}^{1,1} - C_{tq}^1)\,\sigma_S^{AA} + (C_{Qq}^{1,1} + C_{tq}^1)^2\sigma_S^{VV,VV} + (C_{Qq}^{1,1} - C_{tq}^1)^2 \sigma_S^{AA,AA}\,,\\\nonumber
    A_E^{(\text{opt})} & = \big(\sigma_A^{\rm SM} + (C_{Qq}^{1,1} + C_{tq}^1)\,\sigma_A^{VV} + (|C_{Qq}^{1,1}|^2 - |C_{tq}^1|^2)\,\sigma_A^{VV,AA}\big)/\sigma_{t\bar t j}^{(\text{opt})}\,.
\end{align}
For comparison, the rapidity asymmetry in $t\bar t$ production depends on $C_{Qq}^{1,1}$ and $C_{tq}^1$ as
\begin{align}\label{eq:ay-smeft}
    A_{|y|} & = \frac{\sigma_A^\text{SM} + \big(|C_{Qq}^{1,1}|^2 - |C_{tq}^1|^2 \big)\sigma_A^{VV,AA}}{\sigma_S^\text{SM} +  (|C_{Qq}^{1,1}|^2 + |C_{tq}^1|^2)\,\sigma_S^{V+A} + C_{Qq}^{1,1}C_{tq}^1\,\sigma_S^{V-A}}\,.
\end{align}
The cross section yields a constraint on the coefficients in any direction of the two-dimensional parameter space. On the contrary, the charge asymmetries do not. Both $A_E$ and $A_{|y|}$ have a blind direction along $C_{Qq}^{1,1} = - C_{tq}^1$, as long as $\sigma_A^{\rm SM}$ can be neglected. For the rapidity asymmetry, $\sigma_A^{\rm SM}$ is first induced at NLO QCD and very small indeed, see Eq.~\eqref{eq:ay-exp-th}. For the energy asymmetry, $\sigma_A^{\rm SM}$ occurs at LO QCD. It can be sizeable for central jet emission, see Figure~\ref{fig:fixedOrder}, thus breaking the blind direction. In the presence of color-octet operators, the blind direction in $A_E^{\rm (opt)}$ is also broken by axial-vector interference
\begin{align}\label{eq:color-8}
C_{AA}^{q,8} C_{AA}^{q,1} \sim (C_{Qq}^{1,8}-C_{tq}^8)(C_{Qq}^{1,1}-C_{tq}^1)\,.
\end{align}
The rapidity asymmetry does not receive contributions from singlet-octet interference at leading order in QCD, so that the blind direction is broken only at NLO. In Section~\ref{sec:bounds} we will discuss these aspects numerically.

The second pair $\{O_{Qq}^{1,1},\ O_{tu}^1\} \sim \{LL,RR\}$ from Eq.~\eqref{eq:op-set} leads to a simpler geometric structure, because operators with different light-quark chiralities do not interfere with each other in the limit of massless partons. Neglecting down-quark and sea-quark contributions, we obtain
\begin{align}\label{eq:ae-2}
       \sigma_{t\bar t j}^{(\text{opt})} & = \sigma_S^{\rm SM} + (C_{Qq}^{1,1} - C_{tu}^1)\,\sigma_S^{AA} + (|C_{Qq}^{1,1}|^2 + |C_{tu}^1|^2)\,\big(\sigma_S^{VV,VV} + \sigma_S^{AA,AA}\big)\,,\\\nonumber
    A_E^{(\text{opt})} & = \big(\sigma_A^{\rm SM} + (C_{Qq}^{1,1} + C_{tu}^1)\,\sigma_A^{VV} + (|C_{Qq}^{1,1}|^2 + |C_{tu}^1|^2)\,\sigma_A^{VV,AA}\big)/\sigma_{t\bar t j}^{(\text{opt})}\,,
\end{align}
and the asymmetry in $t\bar t$ production reads
\begin{align}\label{eq:ay-2}
    A_{|y|} & = \frac{\sigma_A^\text{SM} + \big(|C_{Qq}^{1,1}|^2 + |C_{tu}^1|^2 \big)\sigma_A^{VV,AA}}{\sigma_S^\text{SM} +  (|C_{Qq}^{1,1}|^2 + |C_{tu}^1|^2)\,\sigma_S^{V+A}}\,.
\end{align}
In this case the operators do not interfere and contribute with the same sign to the observables at the quadratic level. A global fit will set bounds on each individual coefficient and not leave unconstrained directions in the two-dimensional parameter space.

For our numerical analysis, we compute predictions of the $t\bar{t}j$ observables in SMEFT with {\tt Madgraph5\_aMC@NLO 3.0.1}, using the {\tt dim6top} UFO model~\cite{AguilarSaavedra:2018nen} at LO QCD with the same parameters as in the parton-level analysis in Section~\ref{sec:lhc_parton}. In particular, the top quarks are kept stable. The cross section contributions $\sigma_{S,A}^k$ and $\sigma_{S,A}^{kl}$ in Eq.~\eqref{eq:obs-smeft} have been extracted from our simulations, allowing us to obtain $t\bar t j$ observables for arbitrary values of Wilson coefficients. The SM cross sections $\sigma_A^{\text{SM}}$ and $\sigma_S^{\text{SM}}$ are obtained from the NLO parton-level simulations in Section~\ref{sec:lhc_parton}. The statistical Monte-Carlo uncertainty on the energy asymmetry is calculated via error propagation. For the scale uncertainties we calculate the envelope from scale variations.

We also compare our predictions with the rapidity asymmetry $A_{|y|}$ in inclusive $t\bar{t}$ production. For the Standard Model, we use precision predictions at NNLO QCD, including electroweak corrections at NLO~\cite{Czakon:2017lgo}, see Eq.~\eqref{eq:ay-exp-th}. The SMEFT contributions to $A_{|y|}$ are computed at NLO QCD with {\tt Madgraph5\_aMC@NLO}. In our fit to LHC data, we use the latest inclusive measurement of $A_{|y|}$ at 13 TeV~\cite{ATLAS-CONF-2019-026}, see Eq.~\eqref{eq:ay-exp-th}, to derive bounds on the Wilson coefficients.

To give a first idea of the numerical contributions of operators to the observables, we show the effect of one single Wilson coefficient $C_{Qq}^{1,1}$, neglecting all other coefficients. Simultaneous contributions of two operators will be discussed in Section~\ref{sec:bounds}. In Figure~\ref{fig:asy-Qq11} we present predictions of the optimized cross section $\sigma_{t\bar t j}^{\rm opt}$ in the boosted phase-space regime (left) and the charge asymmetries $A_E^{\rm opt}$ and $A_{|y|}$ (right) as functions of $C_{Qq}^{1,1}$.
\begin{figure}[!t]
	\centering
	\includegraphics[width=.49\textwidth]{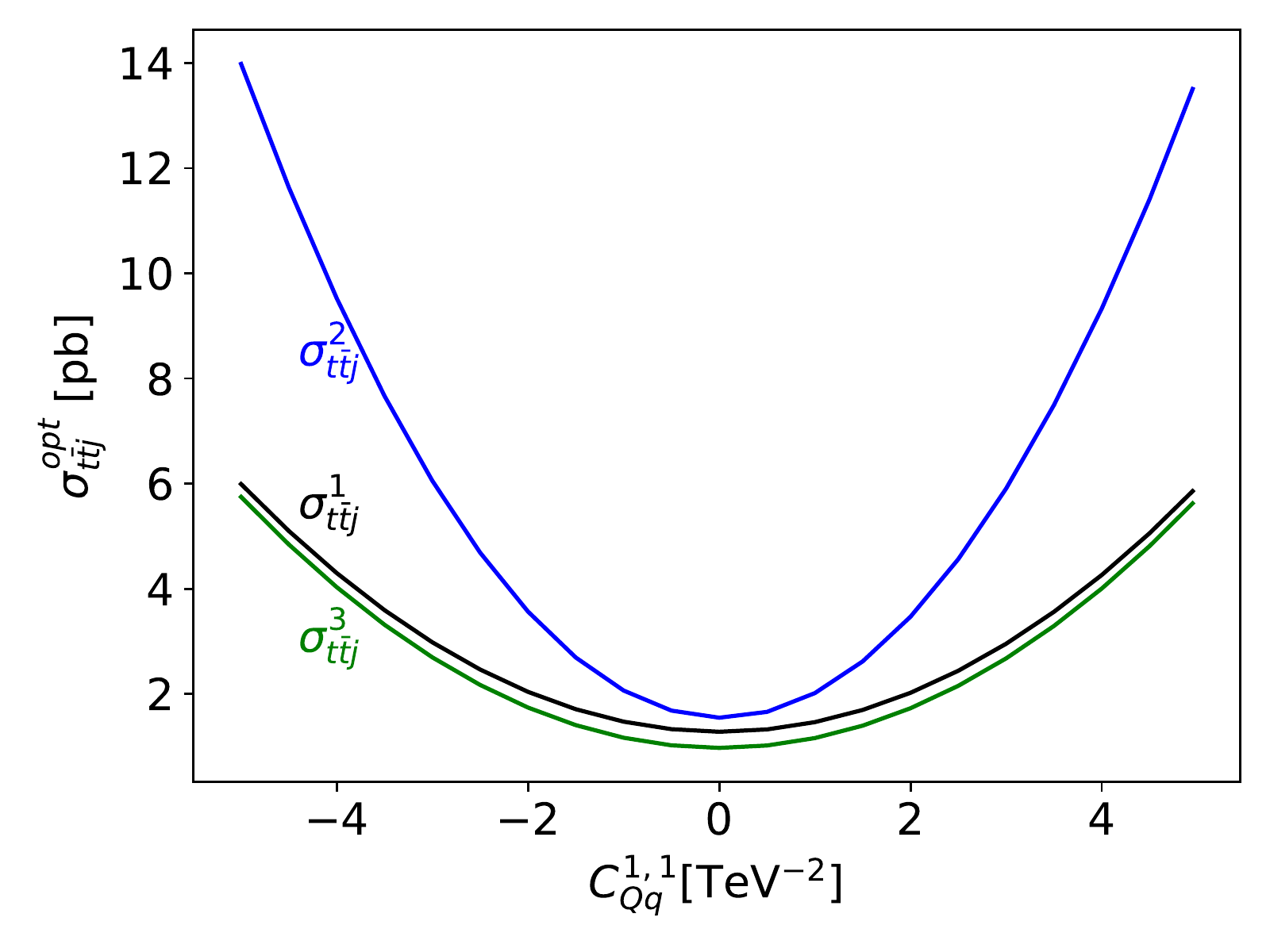}%
	\hspace*{0.3cm}
	\includegraphics[width=.49\textwidth]{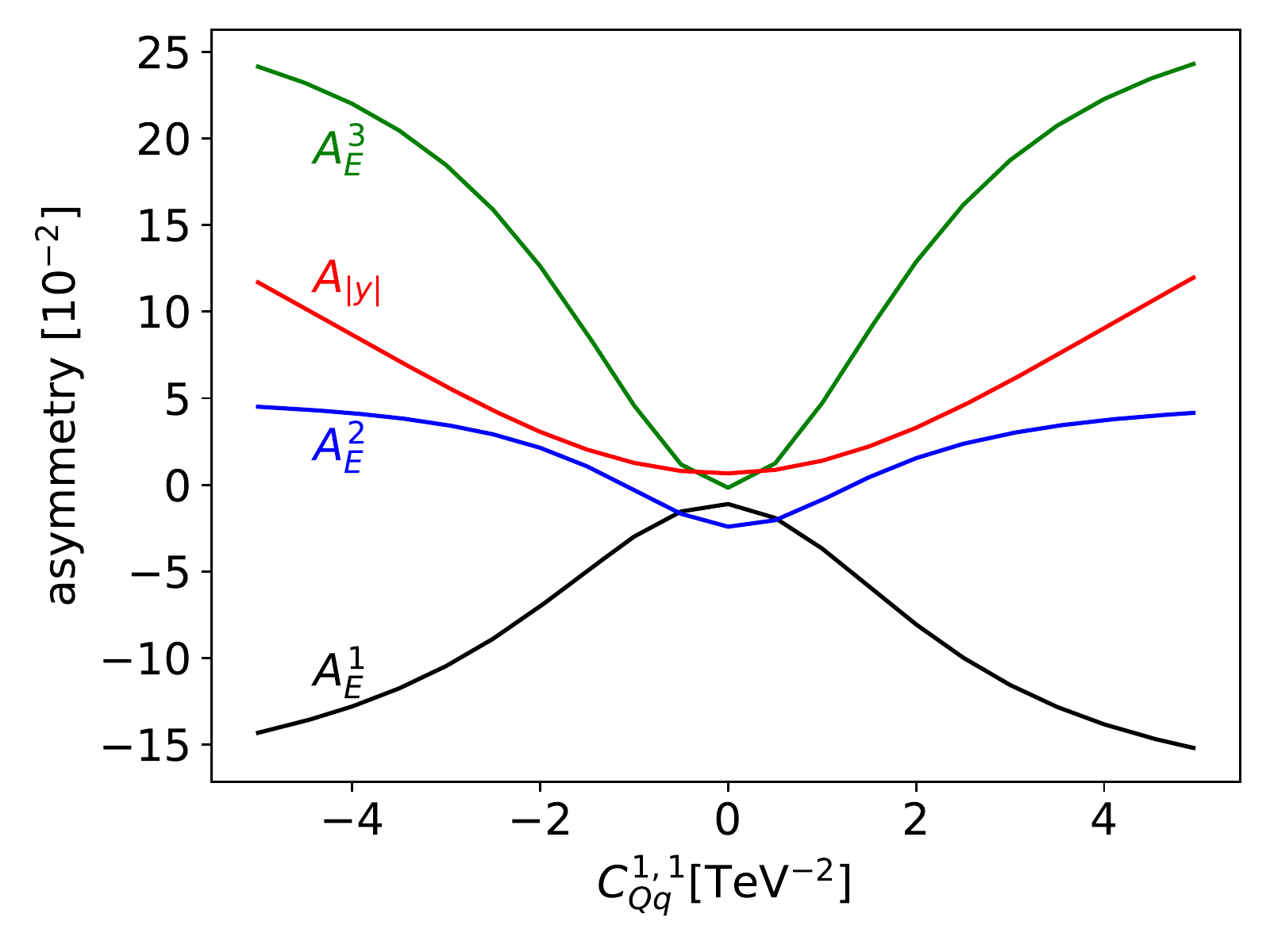}	
	\caption{Dependence of the optimized cross section $\sigma_{t\bar t j}^{\rm opt}$ (left) and the asymmetries $A_E^{\rm opt}$ and $A_{|y|}$ (right) on the Wilson coefficient $C_{Qq}^{1,1}$. For the $t\bar{t} j$ observables, we show predictions in three bins of the jet angle $\theta_j$ as defined in Eq.~\eqref{eq:aebins}.}
	\label{fig:asy-Qq11}
\end{figure}
 The $t\bar t j$ observables are shown for the three bins in $\theta_j$ defined in Eq.~\eqref{eq:aebins}. The cross section in all bins grows quadratically with $C_{Qq}^{1,1}$, as expected from Eq.~\eqref{eq:obs-in-smeft}. The energy asymmetry saturates at large coefficients, where the effect cancels between the numerator and denominator. Contributions to $A_E^{\rm opt}$ are largely symmetric around $C=0$, which means that the interference $\sigma_A^{VV}$ with QCD is small, see Eq.~\eqref{eq:obs-in-smeft}. Notice that the dependence on $C_{Qq}^{1,1}$ is different for each of the three bins $A_E^i$, indicating that $\sigma_A^{VV,AA}$ is positive in bins 2 and 3, but negative in bin 1.
 
 For comparison, we also show the rapidity asymmetry in $t\bar t$ production, as in Eq.~\eqref{eq:ay-smeft} with $C_{tq}^1 = 0$. Since the relative contribution $\sigma_S^{V+A}/\sigma_S^{\rm SM}$ in $t\bar t$ production is much smaller than $\sigma_S^{VV,VV}/\sigma_S^{\rm SM}$ in $t\bar t j$ production, $A_{|y|}$ becomes independent of $C_{Qq}^{1,1}$ only for very large Wilson coefficients beyond the range shown in the figure.

\subsection{Projected LHC sensitivity to effective operators}\label{sec:bounds}
\noindent Using our theory predictions for the cross section and asymmetries from Table~\ref{tab:fixedOrder}, as well as the expected experimental uncertainties from Table~\ref{tab:ParticleLevelExpectedMeasurement}, we predict the sensitivity of an LHC measurement to the Wilson coefficients in SMEFT. To obtain the expected experimental uncertainty on the cross section at parton level, we rescale our particle-level estimates from Table~\ref{tab:ParticleLevelExpectedMeasurement} by the efficiency loss between parton and particle level due to the reduced phase space. We do not make such an adjustment for the energy asymmetry, where the central values at particle and parton level are consistent within uncertainties. All results shown in this section correspond to an LHC measurement of the $t\bar t j$ cross section and optimized energy asymmetry in the selection region called ``boosted'' in Table~\ref{tab:fixedOrder}, based on the Run-2 data set of $139\,\text{fb}^{-1}$.

Assuming that the central value of the measurement will match the SM prediction, we derive the expected confidence limits on the Wilson coefficients $C_i$ from a $\chi^2$ fit with
\begin{equation}
\chi^2 = (x^{\text{SM}} - x^{\text{EFT}})^\top \text{Cov}^{-1} (x^{\text{SM}} - x^{\text{EFT}})\,.
\end{equation}
For a fit of a single observable, $x$ denotes the cross section $\sigma_{t\bar t j}$ in the boosted regime or the optimized energy asymmetry $A_E^{\text{opt}}$ in a specific \thetaj bin. In a combined fit of the asymmetry in all three bins, $x = (A_E^1,A_E^2,A_E^3)^\top$ is a vector. The covariance matrix
\begin{equation}
\text{Cov} = \text{Cov}^{\text{exp. stat.}} + \text{Cov}^{\text{exp. syst.}} + \text{Cov}^{\text{bkg. syst.}} + \text{Cov}^{\text{MC stat.}} + \text{Cov}^{\text{scale}}
\end{equation}
 contains the experimental measurement uncertainties from Table~\ref{tab:ParticleLevelExpectedMeasurement}, as well as the theory uncertainties from Monte-Carlo statistics and scale dependence on the prediction from Table~\ref{tab:fixedOrder}. Scale uncertainties on the SMEFT contributions are assumed to be the same as for the SM prediction. The theory uncertainties on the energy asymmetry in different $\theta_j$ bins are assumed to be uncorrelated. The systematic uncertainties on the background are assumed to be fully correlated or anti-correlated between different bins, depending on the sign of the asymmetry.
 
\begin{figure}[!htbp]
\begin{center}
	\includegraphics[width=.49\textwidth]{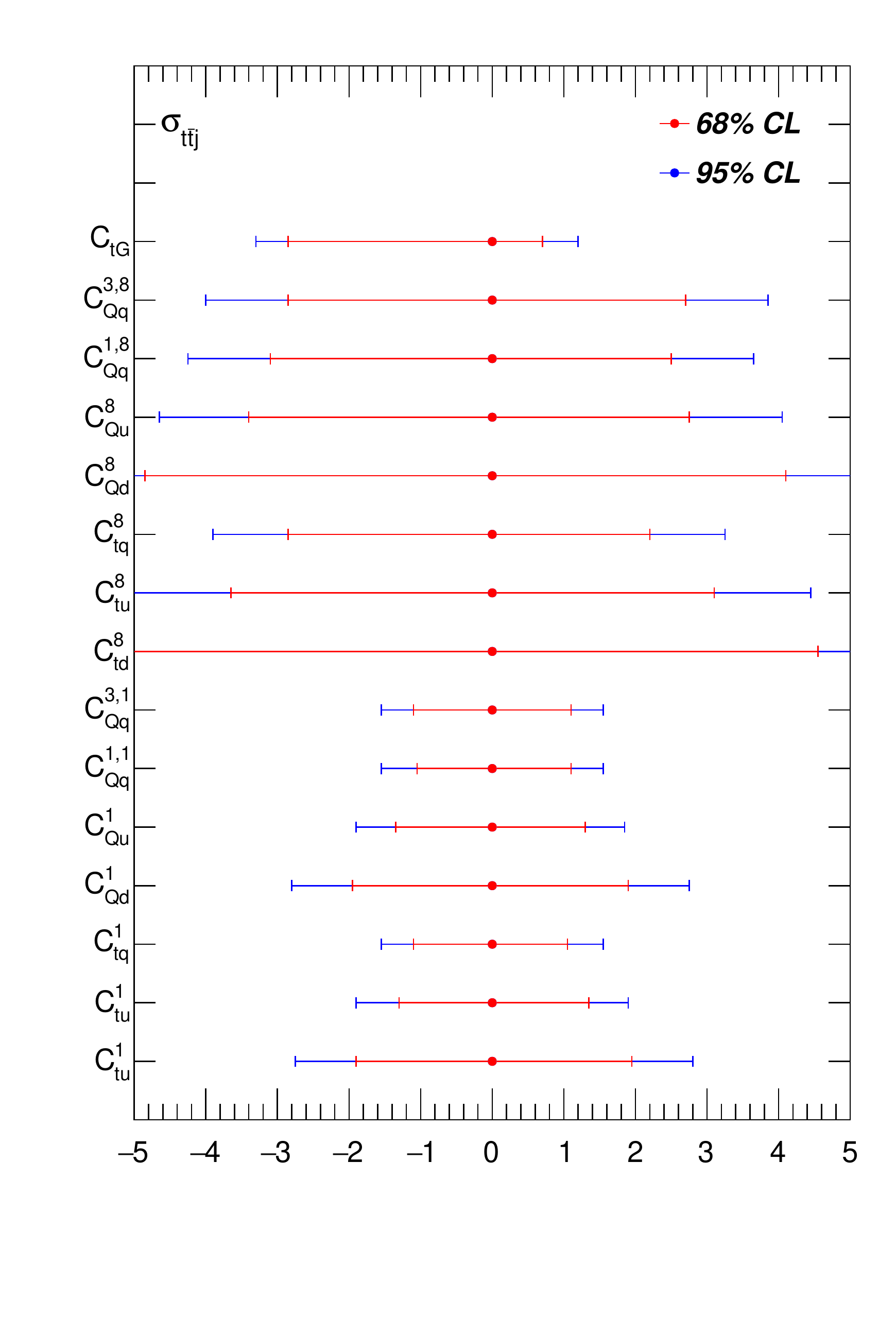}\hspace*{0.3cm}
	\includegraphics[width=.49\textwidth]{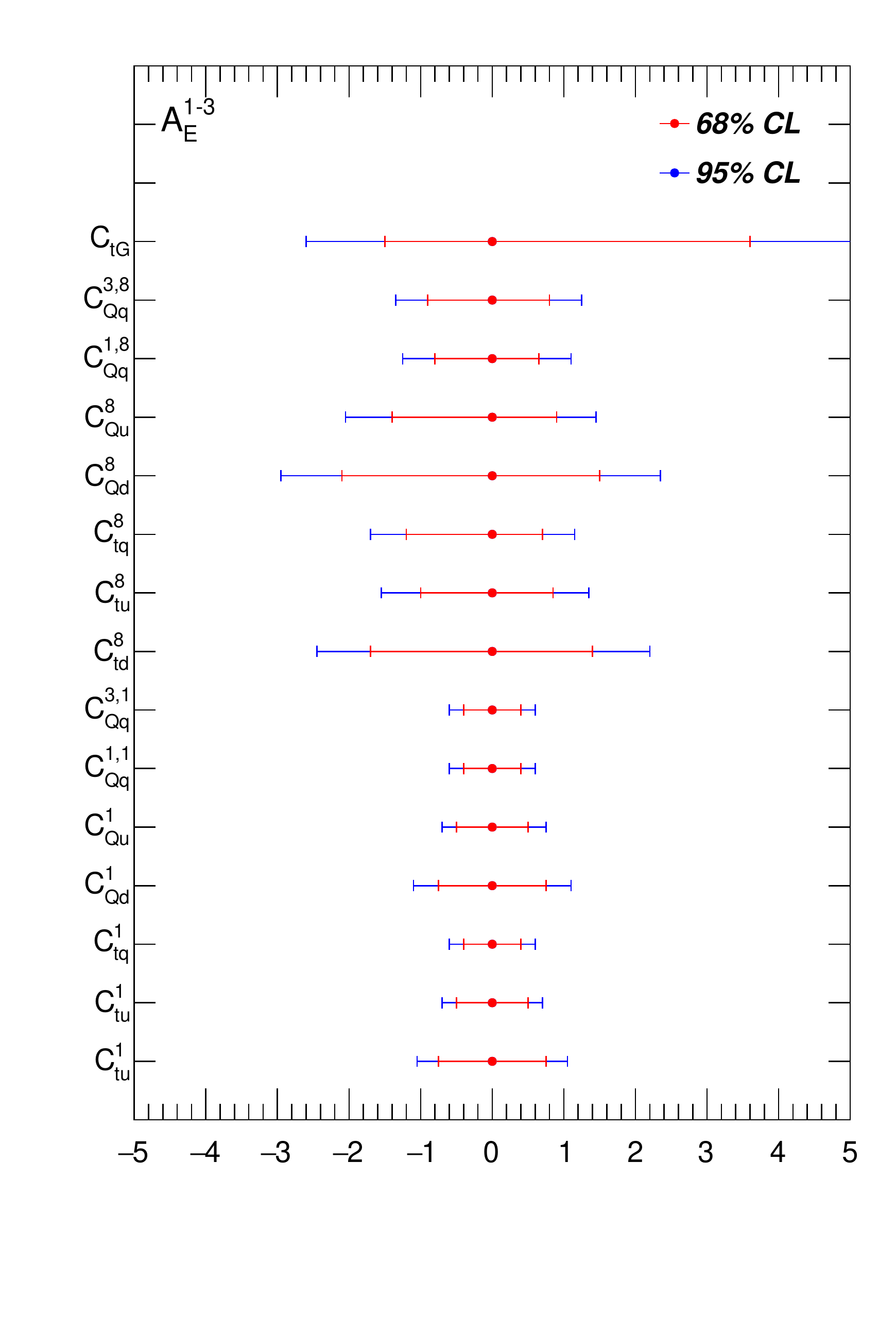} \\\vspace*{-0.5cm}
	\includegraphics[width=.49\textwidth]{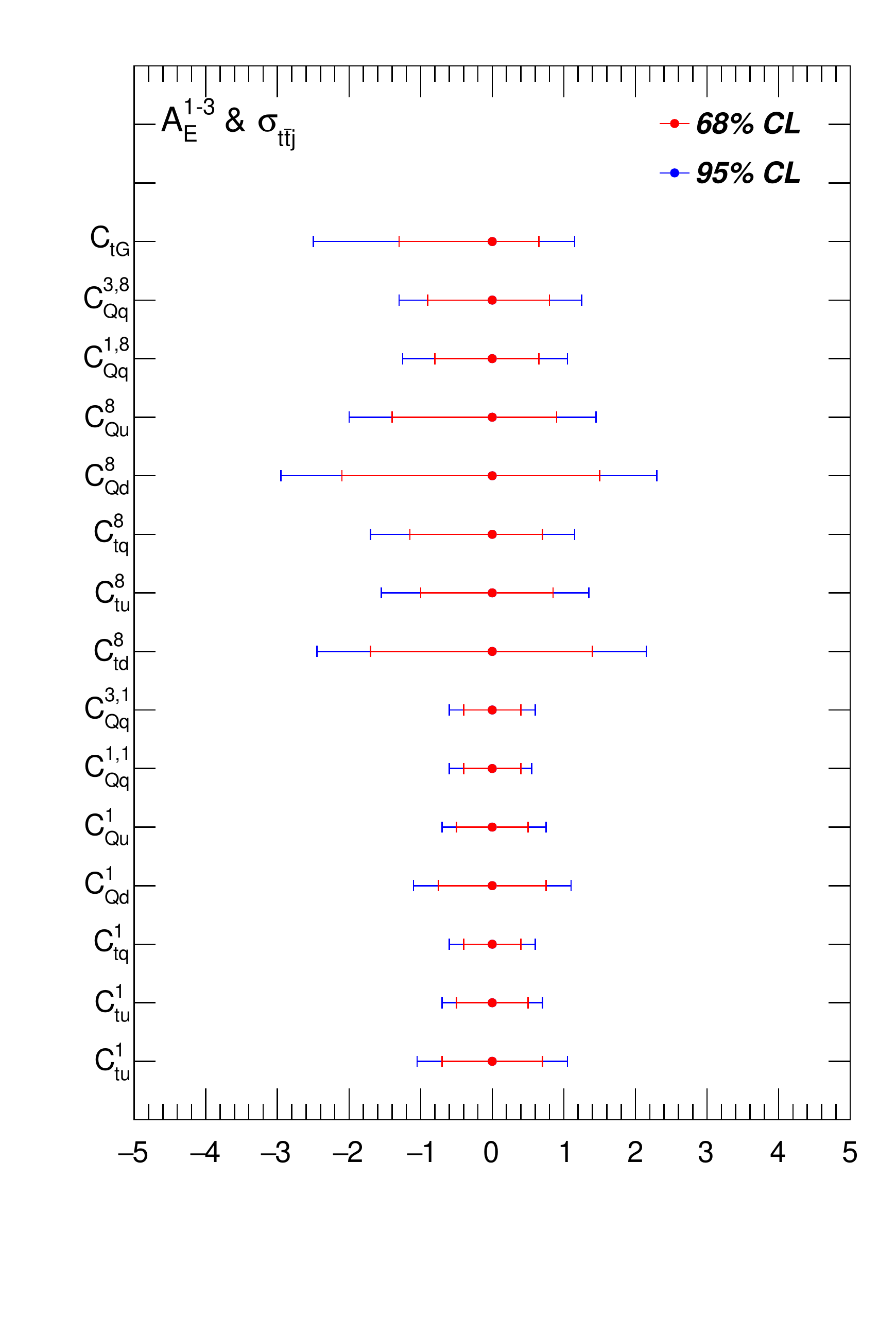}\hspace*{0.3cm}
	\includegraphics[width=.49\textwidth]{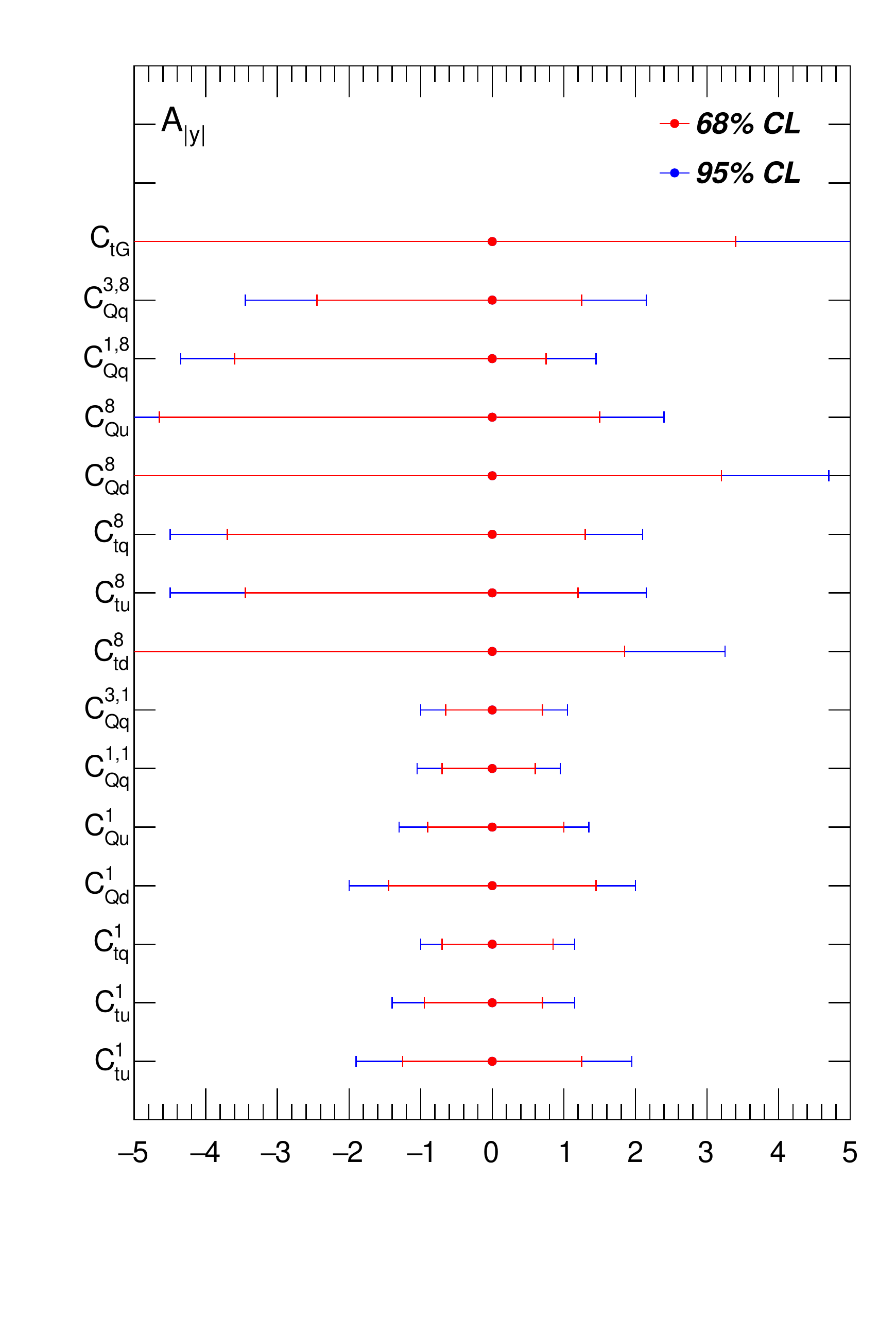}
\end{center}
	\vspace*{-1cm}
\caption{\label{fig:indiv_combi}Expected bounds on individual Wilson coefficients from LHC Run-2 measurements of the cross section $\sigma_{t\bar t j}$ in the boosted regime (top left), the combination of optimized energy asymmetries $A_E^1$, $A_E^2$, $A_E^3$ in all three $\theta_j$ bins (top right), and a combination of these four observables (bottom left). For comparison, we show existing bounds from the rapidity asymmetry in $t\bar t$ production as measured during Run 2~\cite{ATLAS-CONF-2019-026} (bottom right). The bounds on $C_i$ are reported in units of TeV$^{-2}$.}
\end{figure}
In a first approach, we determine the expected bounds on individual Wilson coefficients by including one operator contribution at a time in our fit. Our results are shown in Figure~\ref{fig:indiv_combi} for fits of the cross section $\sigma_{t\bar t j}$ in the boosted regime (top left), a combination of the optimized energy asymmetries in the three $\theta_j$ bins $A_E^1$, $A_E^2$, $A_E^3$ (top right), and a combination of $A_E^{1-3}$ and $\sigma_{t\bar t j}$ (bottom left). The corresponding numerical values for the 68\% and 95\% CL bounds are reported in Tables~\ref{tab:68bounds} and~\ref{tab:95bounds} in the appendix.

Comparing the projected bounds from $\sigma_{t\bar t j}$ (top left) and $A_E^{1-3}$ (top right), we see that for all four-quark operator coefficients the energy asymmetry leads to stronger bounds than the cross section. For most of the coefficients, bin 3 has the highest sensitivity and yields the strongest bounds, see Tables~\ref{tab:68bounds} and~\ref{tab:95bounds}. The cross section has a higher sensitivity to the chromomagnetic dipole moment of the top, $C_{tG}$, than the energy asymmetry. Overall we expect that the combination of $A_E^{1-3}$ and $\sigma_{t\bar t j}$ (bottom left) can give bounds on the Wilson coefficients between $|C| \lesssim 0.5$ and $|C| \lesssim 3$ at 95\% CL, depending on the operator. These expected bounds are comparable to the marginalised bounds obtained from a global fit of $t\bar t$ and single-top observables, as well as $t\bar t W$ and $t\bar t Z$ cross sections~\cite{Brivio:2019ius}, which involves 22 operators with tops. Our findings are promising, as they demonstrate that $t\bar tj$ observables and especially the energy asymmetry can provide additional sensitivity in global SMEFT fits.

For comparison, in Figure~\ref{fig:indiv_combi} we also show bounds on individual Wilson coefficients obtained from the latest Run-2 LHC measurement of the rapidity asymmetry in $t\bar t$ production~\cite{ATLAS-CONF-2019-026} (bottom right). The numerical inputs for the SM prediction and the measurement of $A_{|y|}$ used in the fit are quoted in Eq.~\eqref{eq:ay-exp-th}. Overall the obtained bounds are looser than the predictions for the energy asymmetry (top right), especially for color-octet operators. Notice that stronger bounds from the rapidity asymmetry could be obtained from differential distributions of $A_{|y|}$ as a function of the top-antitop invariant mass and/or rapidity difference $\Delta|y|$.

With a larger data set collected during Run 3 of the LHC or at the HL-LHC, we expect that the sensitivity of the energy asymmetry to SMEFT coefficients will increase. Since the experimental uncertainty is dominated by limited statistics (see Section~\ref{sec:lhc-sm}), we expect the asymmetry-related bounds in Figure~\ref{fig:indiv_combi} to get stronger in case the SM prediction is measured. In turn, the predicted Run-2 sensitivity of the $t\bar t j$ cross section is limited by systematic uncertainties and scale uncertainties. An increased sensitivity will thus relies on improved predictions and reduced systematics.\\

\noindent Bounds on individual Wilson coefficients are useful to explore the relative sensitivity of different observables to the effective operators. However, by considering only one operator at a time we ignore possible degeneracies of operator contributions to an observable, which affect the results of a global fit and can lead to blind directions in the SMEFT parameter space. A global fit of the entire top sector including the energy asymmetry is beyond the scope of this work. Instead we focus on the potential of the energy asymmetry to resolve blind directions occurring in fits of the $t\bar t j$ cross section or the rapidity asymmetry in $t\bar t$ production.

In Figure~\ref{fig:2op-fits} we present the projected bounds for several two-parameter fits, where in each case two four-quark operators are included and all other operator coefficients are set to zero. We have chosen pairs of operators such that we can investigate the effects of the color structure and the quark chirality independently: The top row shows color-singlet operators with different quark chiralities. The operator pairs correspond to the two scenarios discussed in Section~\ref{sec:four-quark}. The middle row shows the same chirality scenarios, but for color-octet operators. The bottom row shows color-singlet versus color-octet operators with the same quark chiralities. Shown are separate fits to the cross section $\sigma_{\ttbar j}$ in the boosted regime (in black), a combination of the optimized energy asymmetries $A_E^{i}$ in three $\theta_j$ bins (in red), and bounds from the rapidity asymmetry $A_{|y|}$ in $\ttbar$ production (in blue).
\begin{figure}[ht!]
\begin{center}
	\includegraphics[width=.49\textwidth]{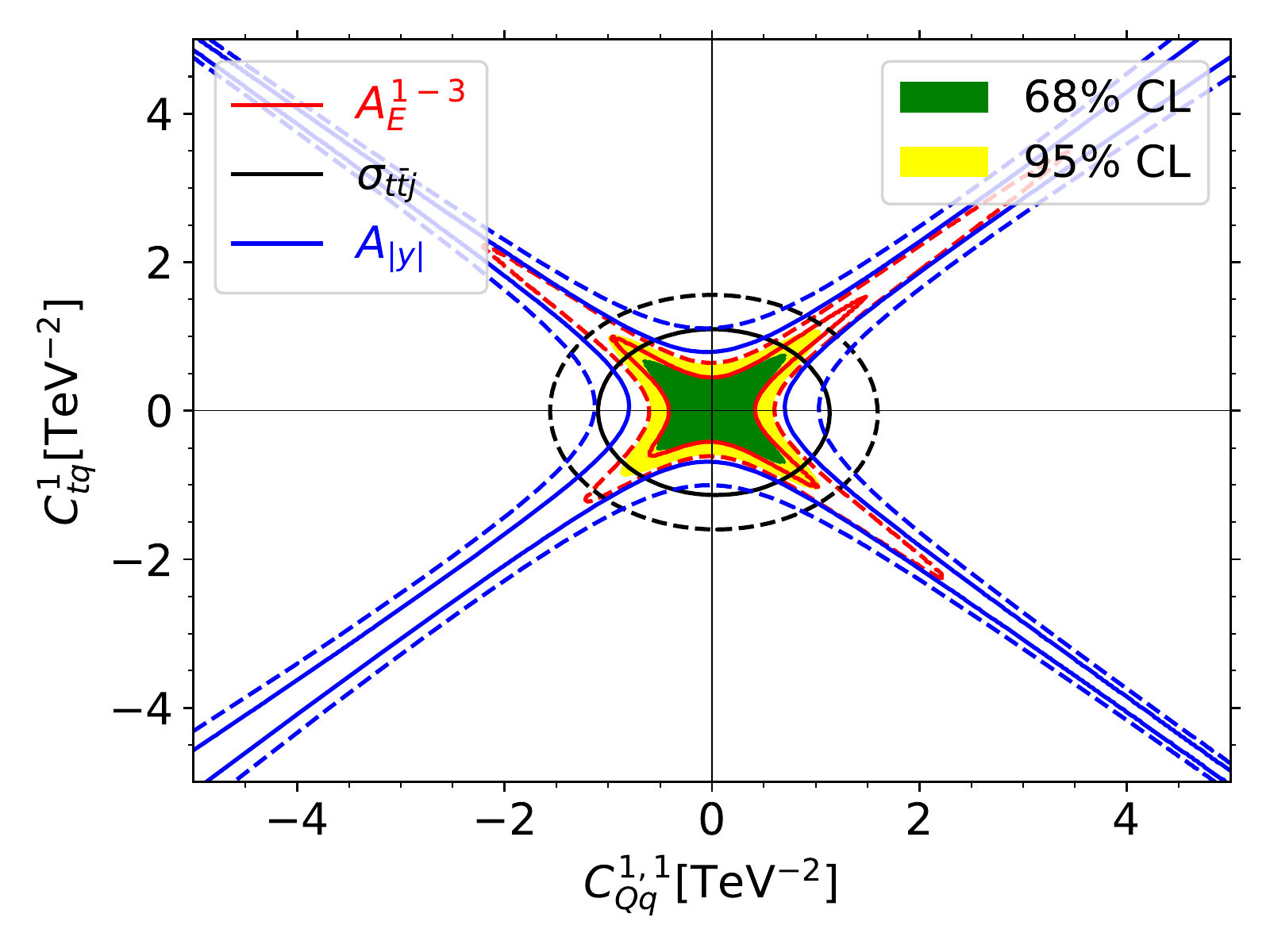}
	\includegraphics[width=.49\textwidth]{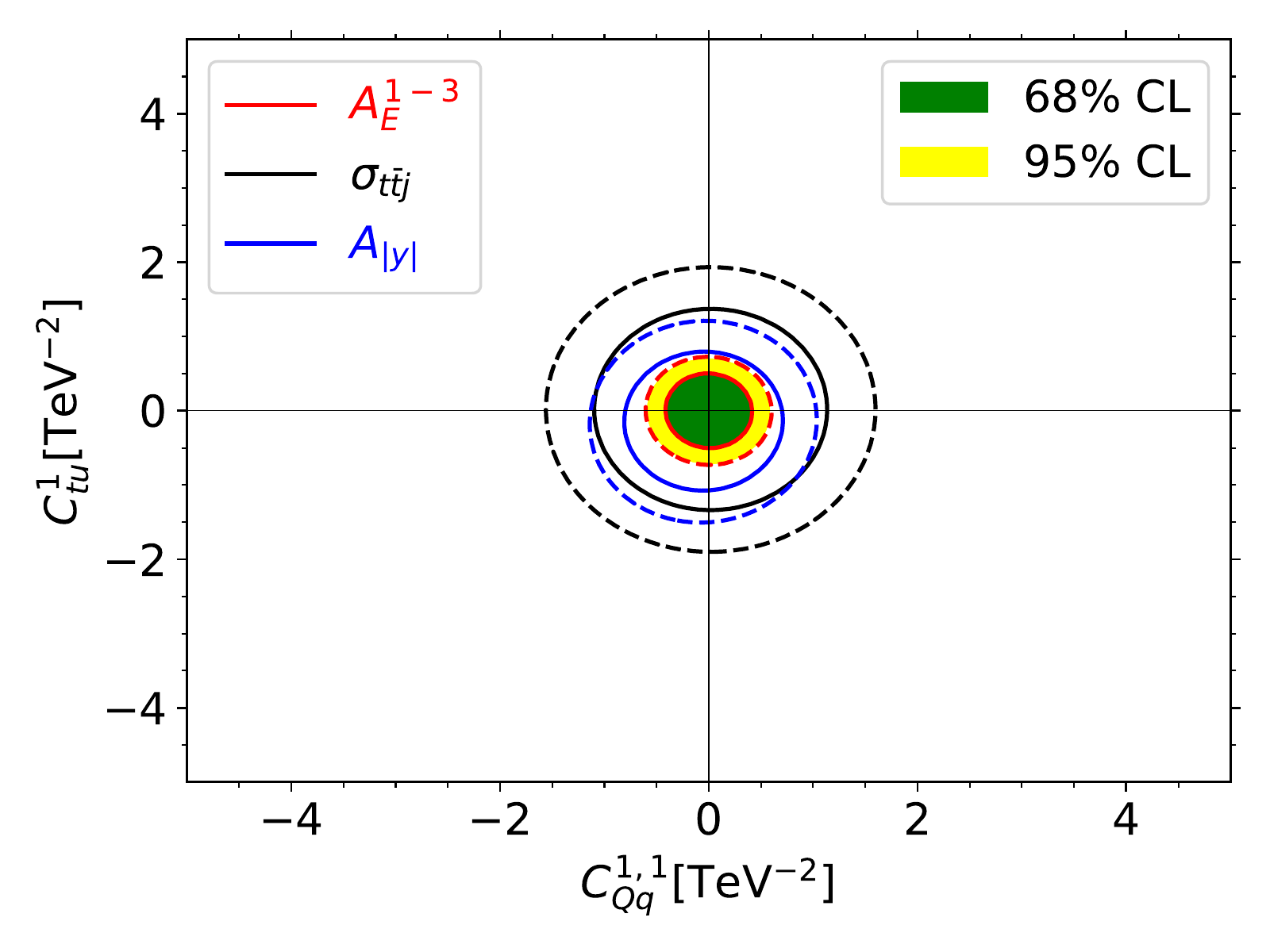} \\
	\includegraphics[width=.49\textwidth]{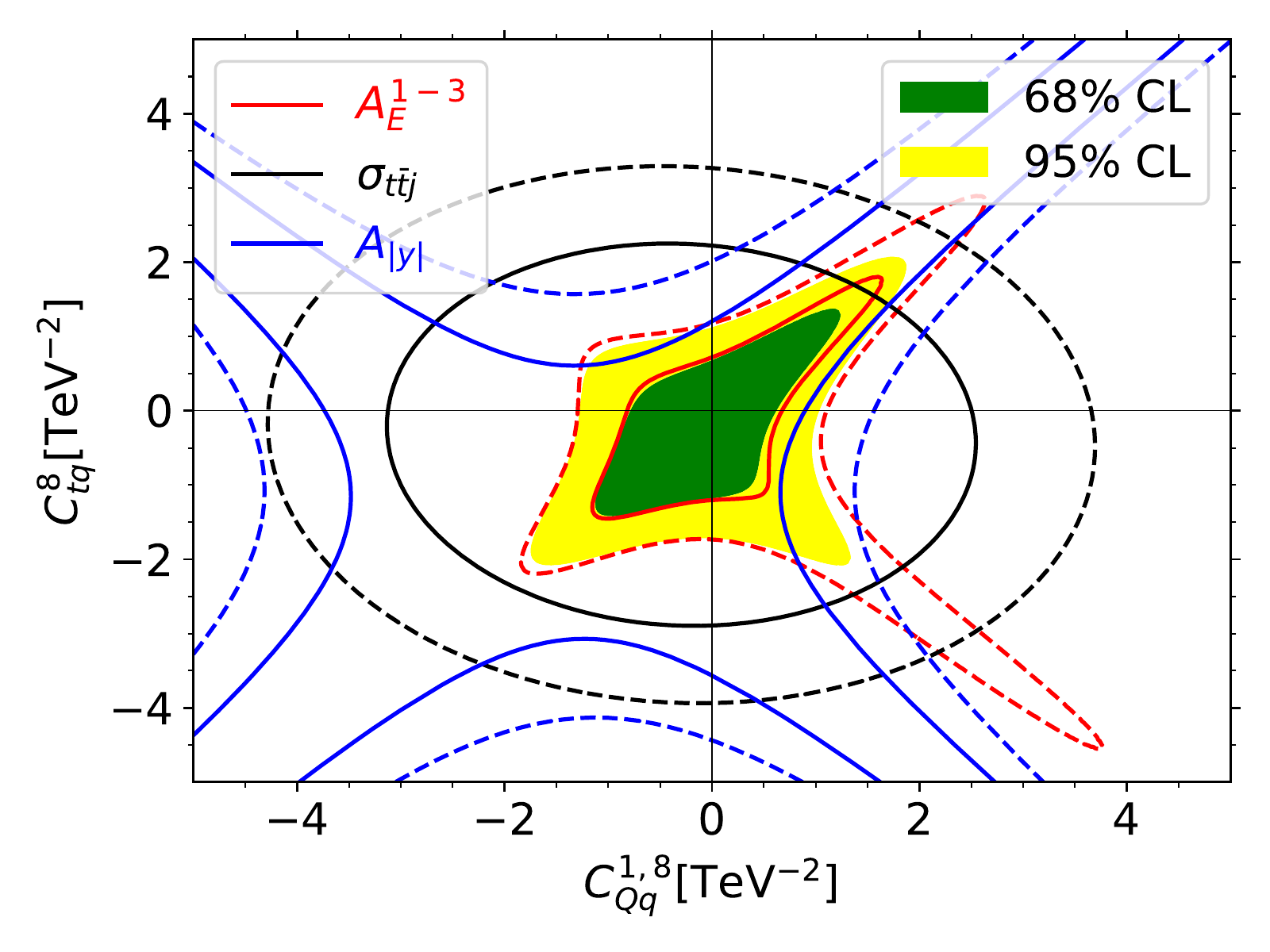}
	\includegraphics[width=.49\textwidth]{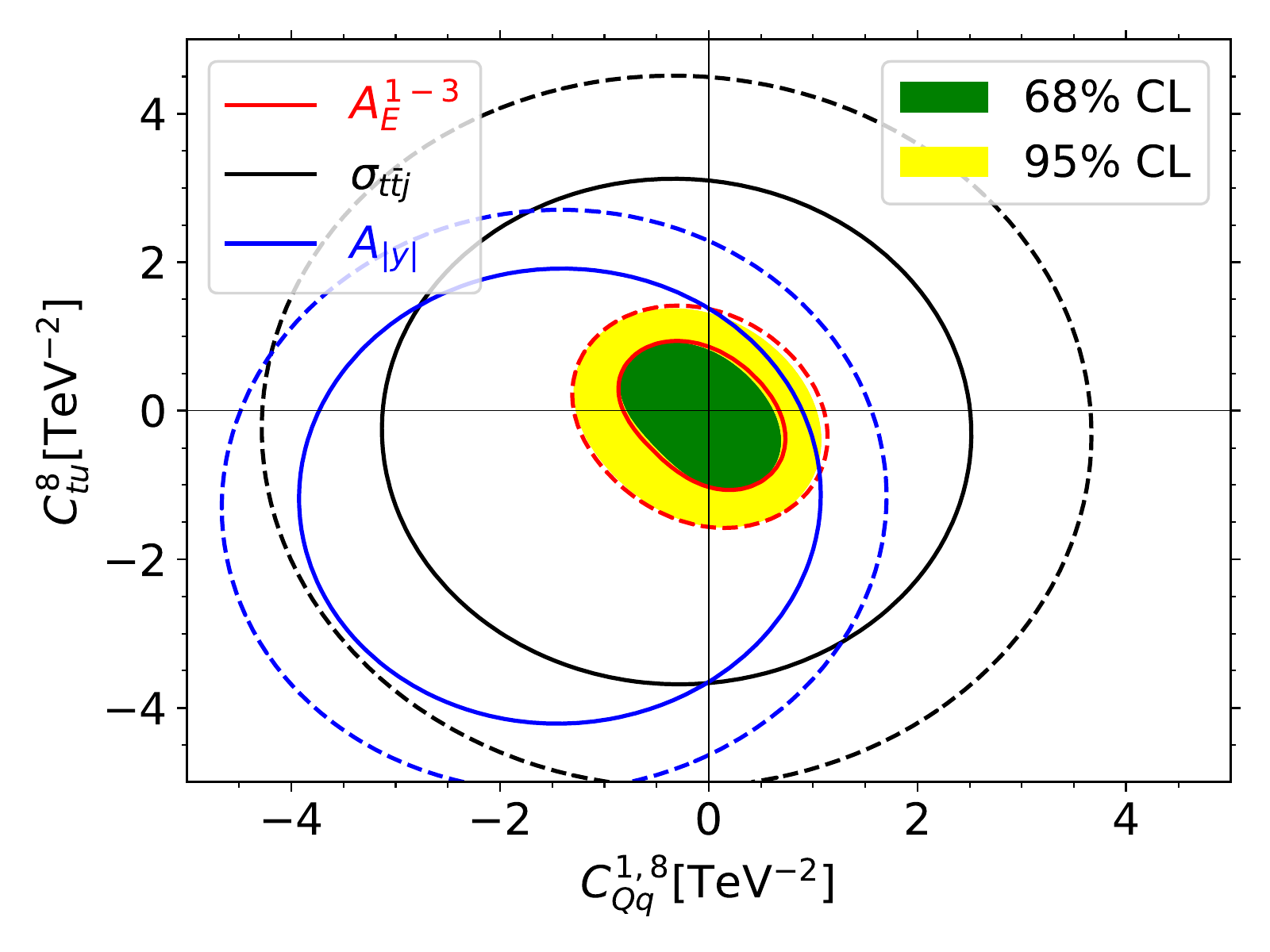} \\
	\includegraphics[width=.49\textwidth]{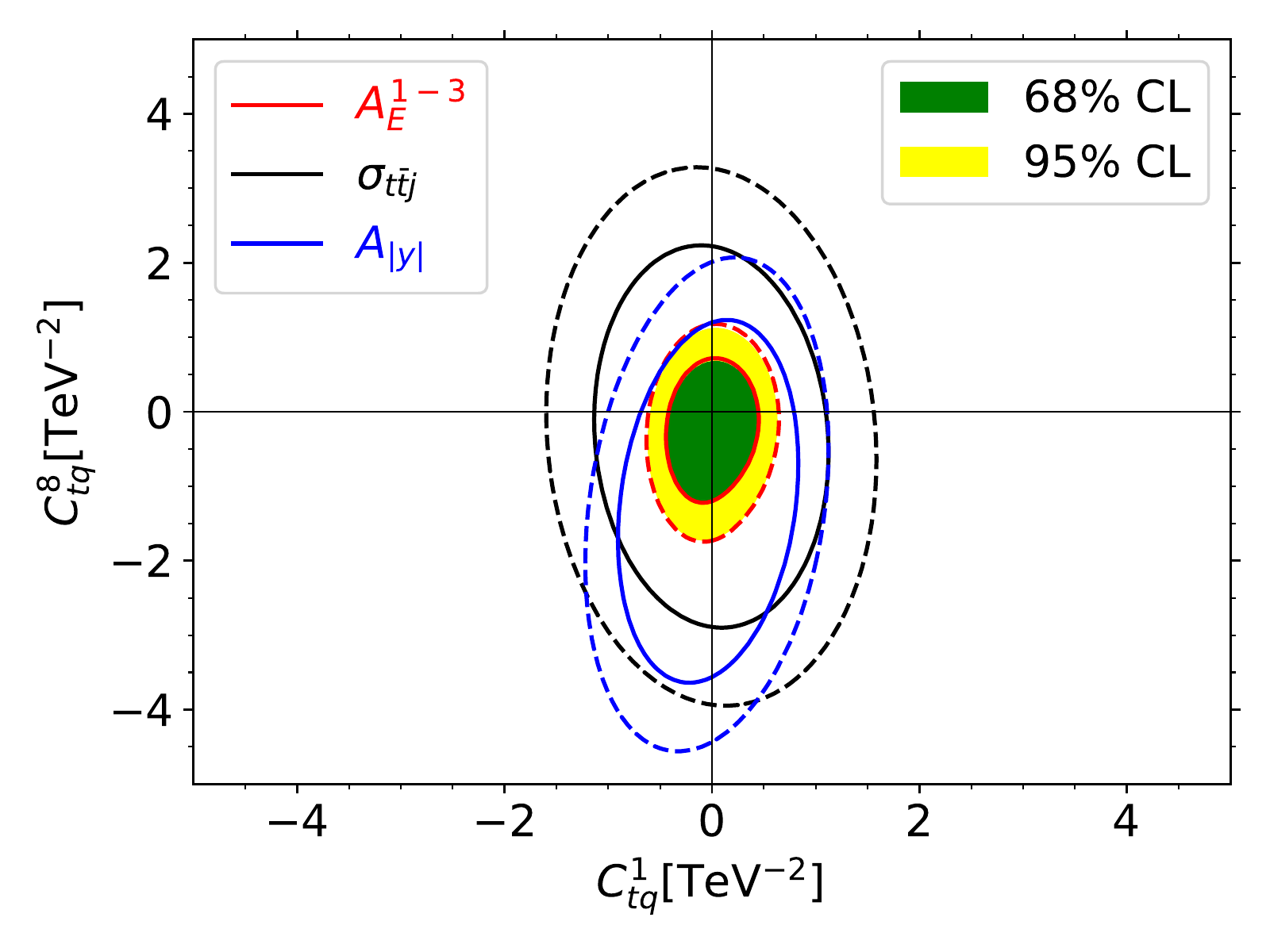} 
	\includegraphics[width=.49\textwidth]{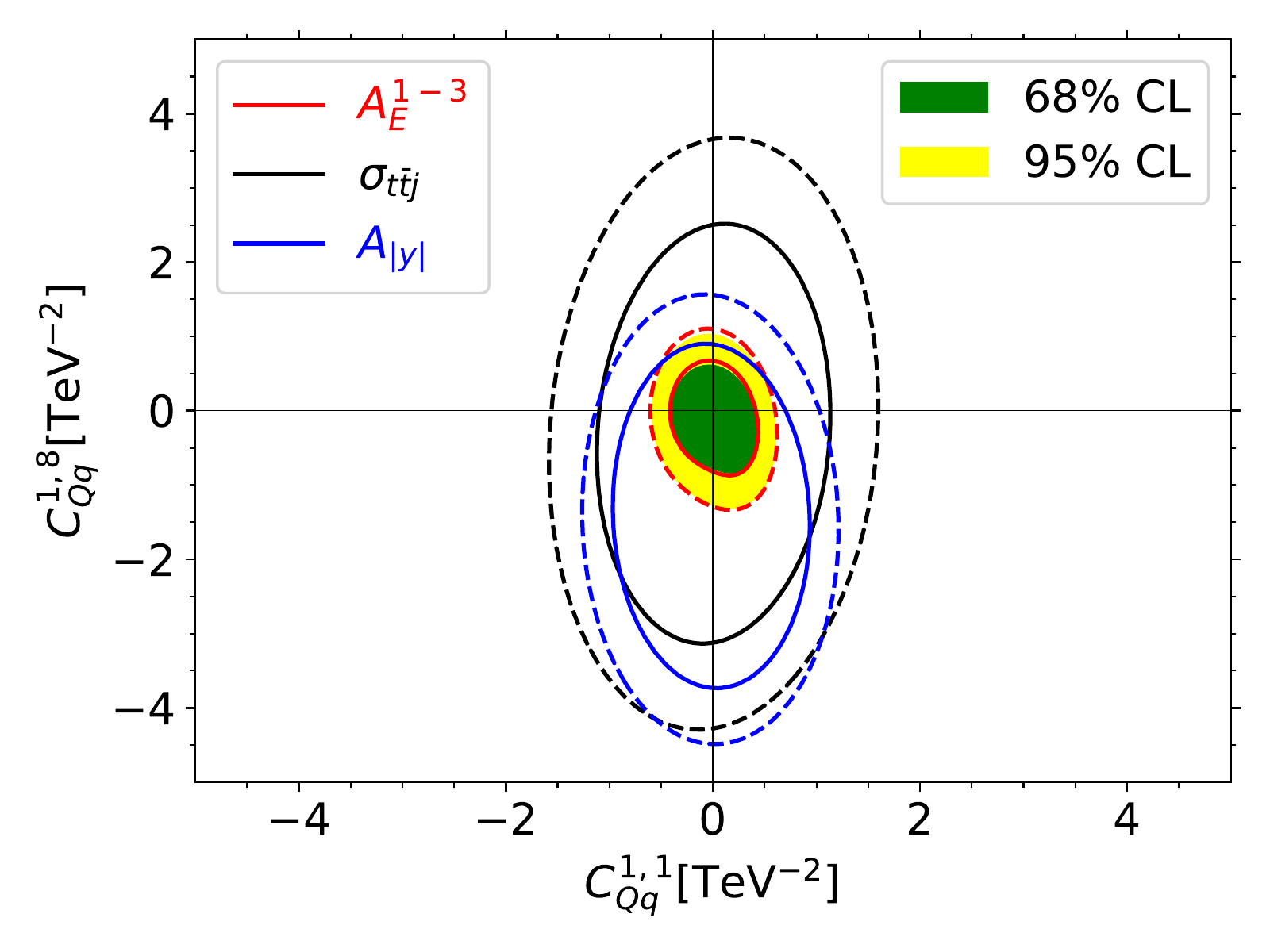}
\end{center}
\caption{Expected bounds on Wilson coefficients from two-parameter fits of the energy asymmetry $A_E^{\text{opt}}$ in all three $\theta_j$ bins (red) and the cross section $\sigma_{t\bar t j}$ in the boosted regime (black) to LHC Run-2 data. For comparison, we show existing bounds from the latest Run-2 measurement of the rapidity asymmetry $A_{|y|}$ in $t\bar t$ production~\cite{ATLAS-CONF-2019-026} (blue). Solid and dashed lines mark the 68\% and 95\% confidence levels for each observable. Green and yellow regions show the 68\% and 95\% CL limits of a combined fit to all five observables.\label{fig:2op-fits}}
\end{figure}

 Focusing on the operator pair $\{O_{Qq}^{1,1},Q_{tq}^1\}$ in Figure~\ref{fig:2op-fits}, top left, we see that the rapidity asymmetry $A_{|y|}$ constrains the parameter space in the form of a hyperbola. This confirms our analytic discussion from Section~\ref{sec:four-quark}, where we identified a blind direction along $|C_{Qq}^{1,1}| = |C_{tq}^1|$. For the combination of energy asymmetries $A_E^{1-3}$, we do not encounter such a blind direction, although the bounds along the diagonals are loose. In this combination, $A_E^2$ probes the direction $|C_{Qq}^{1,1}| = |C_{tq}^1|$ through a sizeable SM contribution of $\sigma_A^{\rm SM}$, see Eq.~\eqref{eq:obs-in-smeft}. Since the interference $\sigma_A^{VV}$ is numerically small, the bounds are symmetric around $C=0$. The cross section $\sigma_{\ttbar j}$ constrains the parameter space to an ellipse centered around the origin. As for the asymmetries, this shows that operator contributions of $\mathcal{O}(C^2/\Lambda^4)$ are relevant in setting the bounds.

The combined fit of $\sigma_{t\bar t j}$, $A_E^{1-3}$ and $A_{|y|}$ (green and yellow regions) for $\{O_{Qq}^{1,1},Q_{tq}^1\}$ shows that the energy asymmetry plays an important role, improving both the bounds on individual operators and the sensitivity to the chirality of the top. For the second operator pair $\{O_{Qq}^{1,1},Q_{tu}^1\}$ (top right), both asymmetries constrain the parameter space to an ellipse. The reason is that these operators do not interfere and they contribute with the same sign to the asymmetries, see Eqs.~\eqref{eq:ae-2} and \eqref{eq:ay-2}. This demonstrates that operator interference changes the geometric shape of the asymmetry bounds.

Color-octet operators with the same chiralities as discussed before induce additional contributions at $\mathcal{O}(\Lambda^{-2})$ and $\mathcal{O}(\Lambda^{-4})$. For the operator pair $\{O_{Qq}^{1,8},Q_{tq}^8\}$ (middle left), the energy asymmetry probes the following directions in the two-parameter space
\begin{align}
 C_{Qq}^{1,8} + C_{tq}^8\,,\quad  C_{Qq}^{1,8} - C_{tq}^8\,,\quad (C_{Qq}^{1,8} + C_{tq}^8)^2\,,\quad (C_{Qq}^{1,8} - C_{tq}^8)^2\,,\quad |C_{Qq}^{1,8}|^2 - |C_{tq}^8|^2\,.
\end{align}
The interference of color-octet operators with QCD shifts the bounds in the two-dimensional parameter space. In particular, the combined bound from the energy asymmetry bins is distorted due to a sizeable shift of $A_E^2$. For the operator pair $\{O_{Qq}^{1,8},Q_{tu}^8\}$ (middle right), the energy asymmetry probes schematically the directions
\begin{align}
r_{qg}\,(C_{Qq}^{1,8} + C_{tu}^8) + C_{Qq}^{1,8}\,,\qquad r_{qg}\,(|C_{Qq}^{1,8}|^2 + |C_{tu}^8|^2) + |C_{Qq}^{1,8}|^2\,.
\end{align}
The shape of the bounds looks thus similar as for color-singlet operators. Notice that the expected bounds from $t\bar t j$ observables are dominated by $\mathcal{O}(\Lambda^{-4})$ contributions, while the asymmetry in $t\bar t$ production is more sensitive to contributions of $\mathcal{O}(\Lambda^{-2})$. This explains the shift of the $A_{|y|}$ bounds  in the two-dimensional plane in the presence of color-octet operators.

The bounds on color-singlet operators are generally stronger than for color-octets, as we see by comparing the diagrams in the top and middle rows or the axes of the ellipses in the bottom row. This is due to the QCD structure of the amplitudes, which enhances color singlets. In Table~\ref{tab:tt-ttj-ops} we see that $t\bar t$ production probes the combination $|C^8|^2 + 9/2\, |C^1|^2$, neglecting interference with QCD. In $t\bar t j$ production, we encounter combinations
\begin{align}
|C^8|^2 + \frac{3}{2}\,|C^1|^2\quad \text{and}\quad |C^8|^2 + \frac{4}{3}\,(C^1 C^8 + C^8 C^1)\,.
\end{align}
The emission of a jet changes the relative sensitivity to color-singlet and color-octet operators, which breaks the blind direction along $9/2\,|C^8|^2 - |C^1|^2$ present in $t\bar t$ production.

In summary, the energy asymmetry in $t\bar t j$ production has a high sensitivity to four-quark operators with different top chiralities and color structures. Measuring and including the energy asymmetry in a global SMEFT fit will probe new directions in the parameter space of Wilson coefficients and improve the sensitivity to individual operators. An interesting complementary observable could be the rapidity asymmetry in $t\bar t j$ production, which is also very sensitive to new vector and axial-vector currents~\cite{Berge:2012rc}.

\section{Conclusions and outlook}\label{sec:conclusions}
\noindent The energy asymmetry in $t\bar t j$ production provides a new handle on top quark interactions. In this work we have provided realistic predictions of the energy asymmetry in QCD and in SMEFT for a planned measurement in LHC data. We have computed the energy asymmetry in the Standard Model in a realistic analysis setup based on NLO QCD predictions and including effects of the parton shower and hadronization. Our analysis has been optimized to maximize the asymmetry by applying appropriate phase-space selections of the top-antitop-jet system. Focusing on the final state with one leptonic and one hadronic top, we have obtained particle-level predictions for the optimized energy asymmetry in three bins of the jet angle $\theta_j$. In the most sensitive bin with central jet emission, we find that the energy asymmetry reaches $A_E^2\approx - 2.2$\% in the boosted regime.

In a data set of 139~fb$^{-1}$ from Run 2 the energy asymmetry in QCD can be measured with about $40$\% experimental uncertainty, corresponding to 2.5 standard deviations from zero. Our projections for Run 3 and the HL-LHC show that the measurements can reach an improved accuracy of about $30$\% and down to $10$\%, respectively. The statistical limitations at Run 2 are thus overcome, and the significance is increased to 3 and more than 5 standard deviations, respectively.

Given the promising results of our QCD analysis, we have analyzed for the first time the impact of new top interactions on the energy asymmetry within the SMEFT framework. We have computed the contributions of all relevant top-quark operators to the asymmetry, following the same LHC analysis as for the Standard Model. The energy asymmetry probes a large number of particular combinations of Wilson coefficients and is highly sensitive to axial-vector currents. Based on our SMEFT predictions and our estimates for the experimental uncertainties, we have extracted the expected bounds on individual operator coefficients from a future fit to LHC measurements.

We find that for all four-quark operators with tops the differential measurement of the energy asymmetry in $t\bar t j$ production provides a better sensitivity than the measurement of the inclusive $\sigma_{t\bar t j}$ cross section in the same phase-space region. We have also compared the sensitivity of $t\bar t j$ observables with the recent ATLAS measurement of the inclusive rapidity asymmetry in $t\bar t$ production. A measurement of the energy asymmetry will lead to improved constraints on dimension-6 operators, individually and in combination with existing charge asymmetry measurements. To explore the potential of the energy asymmetry to break blind directions between different operator coefficients, we have performed 2-parameter fits of several operator pairs with different chirality and color structures. We find that the energy asymmetry probes different combinations of Wilson coefficients than the $t\bar t j$ cross section or the rapidity asymmetry. Based on our numerical predictions, we expect that the energy asymmetry can have a significant impact on future global SMEFT fits. 

In conclusion, we advocate a measurement of the energy asymmetry at the LHC, having demonstrated its feasibility with a realistic collider analysis. With our SMEFT analysis, we have shown that such a measurement can play a crucial role in global SMEFT interpretations of observables in the top sector.\\

\begin{center} \textbf{Acknowledgments} \end{center}
\noindent We thank the authors of Ref.~\cite{Czakon:2017lgo} for providing us with the symmetric and asymmetric contributions to the rapidity asymmetry in the SM. The research of SW is supported by the Carl Zeiss foundation through an endowed junior professorship and by the German Research Foundation (DFG) under grant no. 396021762--TRR 257.  AB's work is funded by the Deutsche Forschungsgemeinschaft (DFG, German Research Foundation) under Germany’s Excellence Strategy – EXC 2118 PRISMA$^+$ – 390831469. PB has received funding from the European Union’s Horizon 2020 research and innovation programme under the Marie Sklodowska-Curie grant agreement no. 797520. 

\appendix

\section{Expected uncertainties of an LHC measurement}\label{app:uncertainties}
\noindent Based on our assumptions on event reconstruction from Section~\ref{sec:lhc_particle}, we estimate the expected experimental uncertainty on the energy asymmetry \AEopt at particle level in an LHC measurement. For this purpose, we rewrite the definition of \AEopt from Eq.~\eqref{eq:ea-opt} as
\begin{equation}\label{eq:ea-opt-rewritten}
	A_E^{\rm opt}(\thetaj) = \frac{\sigma^+(\thetaj)-\sigma^-(\thetaj)}{\sigma^+(\thetaj)+\sigma^-(\thetaj)}\,,
\end{equation}
where  
\begin{align}\label{eq:ea-opt-xsec}
\sigma^+(\thetaj) & \equiv \sigma_{\ttjet}(\thetaj,\Delta E > 0,y_{\ttjet}>0) + \sigma_{\ttjet}(\pi-\thetaj,\Delta E > 0,y_{\ttjet}<0)\,,\\\nonumber
\sigma^-(\thetaj) & \equiv \sigma_{\ttjet}(\thetaj,\Delta E < 0,y_{\ttjet}>0) + \sigma_{\ttjet}(\pi-\thetaj,\Delta E < 0,y_{\ttjet}<0)\,.
\end{align}
Our uncertainty estimate applies to a future LHC measurement using a data set of a certain integrated luminosity $L$. To obtain the measured energy asymmetry, we need to extract $\sigma^+$ and $\sigma^-$ from the detected events. We assume selection criteria at detector level to select the events of interest with high purity, as commonly done in LHC measurements of differential cross sections, for instance in Ref.~\cite{CERN-EP-2019-149}. We then correct the selected event numbers with $\Delta E > 0$ and $\Delta E < 0$, called $\Ndata^+$ and $\Ndata^-$, for the corresponding backgrounds $B^+$ and $B^-$ from processes other than \ttbar production and extrapolate the result from the phase space defined by the detector-level selection to the phase space defined by the particle-level selection. In summary, we obtain the cross sections $\sigma^+$ and $\sigma^-$ as
\begin{equation}\label{eq:sigma-meas}
\sigma^+(\thetaj) = \frac{\Ndata^+(\theta_j)-\Nbkg^+(\theta_j)}{L}\,\frac{\effPart^+(\theta_j)}{\effReco^+(\theta_j)}\,,\quad \sigma^-(\thetaj) = \frac{\Ndata^-(\theta_j)-\Nbkg^-(\theta_j)}{L}\,\frac{\effPart^-(\theta_j)}{\effReco^-(\theta_j)}\,.
\end{equation}
Here $\effPart$ is the efficiency of particle-level selection criteria in the phase space defined by the detector-level selection, and $\effReco$ is the efficiency of detector-level selection criteria in the phase space defined by the particle-level selection. As before, the indices $+$ and $-$ refer to events with $\Delta E > 0$ and $\Delta E < 0$, respectively. In Eq.~\eqref{eq:sigma-meas}, we assume that the detector-level objects perfectly match the particle-level objects, {\it i.e.}, that there is no need to correct for detector effects (commonly referred to as \textit{unfolding}). Furthermore we assume that the selection efficiencies and the background do not depend on the sign of $\Delta E$, so that
\begin{align}\label{eq:background}
\effPart^+(\thetaj) & =\effPart^-(\thetaj)\equiv\effPart(\theta_j)\,,\quad \effReco^+(\thetaj)=\effReco^-(\thetaj)\equiv\effReco(\theta_j)\,,\\\nonumber
\Nbkg^+(\thetaj) & =\Nbkg^-(\thetaj)\equiv \frac{\Nbkg(\theta_j)}{2}\,.
\end{align}
These assumptions might not be exactly valid in a real-detector environment and need to be tested in an analysis based on real data. Here we apply these simplifications to obtain an approximate estimate of the expected experimental uncertainties. Under these assumptions, most of the dependence on the selection efficiencies cancels in the normalized asymmetry, and we obtain the optimized energy asymmetry as
\begin{equation}\label{eq:ea-opt-meas}
	A_{E,\text{meas}}^{\rm opt}(\thetaj) = \frac{\Ndata^+(\thetaj)-\Ndata^-(\thetaj)}{\Ndata^+(\thetaj)+\Ndata^-(\thetaj)-\Nbkg(\thetaj)}\,.
	\end{equation}
Based on this formula, we estimate the main sources of experimental uncertainties. The numbers of detected events, $\Ndata^+$ and $\Ndata^-$, are Poisson-distributed with absolute statistical uncertainties $\sqrt{\Ndata^+}$ and $\sqrt{\Ndata^-}$, respectively. From these uncertainties we obtain the overall statistical uncertainty on the energy asymmetry, $\Delta A_{E}^{\rm stat}$, by error propagation. The expected number of background events \Nbkg is affected by a systematic uncertainty due to the imperfect background estimate. We refer to the corresponding background uncertainty of the asymmetry as $\Delta A_E^{\rm bkg}$. We finally assume that all detector-related uncertainties cancel between the numerator and denominator of the asymmetry and neglect them. The expected total absolute uncertainty on the measured energy asymmetry is thus given by
\begin{equation}
\Delta A_{E}^{\rm tot}(\thetaj) = \sqrt{\big( \Delta A_{E}^{\rm stat}(\thetaj)\big) ^2 + \big(\Delta A_E^{\rm bkg}(\thetaj)\big)^2}\,.
\end{equation}
Since the energy asymmetry is relatively small, it is convenient to approximate the event numbers $D^+$ and $D^-$ using Eq.~\eqref{eq:sigma-meas} as
\begin{equation}
\Ndata^+(\thetaj)\approx \Ndata^-(\thetaj) \approx
\frac{\sigmaopt(\thetaj)}{2}\frac{L}{\fttbar(\thetaj)} \frac{\effReco(\thetaj)}{\effPart(\thetaj)}\,,\quad\text{with}\quad \fttbar=\frac{\Ndata^+ + \Ndata^- - \Nbkg}{\Ndata^+ + \Ndata^-}\,,
\end{equation}
wherever it does not affect the derived uncertainty on the asymmetry. Here $\fttbar$ is the fraction of \ttbar events among the selected number of events. Based on this approximation and using error propagation, we obtain the expected statistical and background uncertainties
\begin{align}
\Delta A_E^{\rm stat}(\thetaj) & \approx \sqrt{\frac{1}{L\,\sigmaopt(\thetaj) \fttbar(\thetaj) }\frac{\effPart(\thetaj)}{\effReco(\thetaj)}}\,,\\
\Delta A_E^{\text{bkg}}(\thetaj) & \approx |A_{E,\mathrm{meas}}^{\text{opt}}(\thetaj)| \frac{1-\fttbar(\thetaj)}{\fttbar(\thetaj)}\frac{\Delta\Nbkg}{\Nbkg}\,,
\end{align}
where $\Delta \Nbkg/\Nbkg$ is the relative uncertainty of the background estimate. We observe that the statistical uncertainty scales as $1/\sqrt{L}$, while the background uncertainty does not depend on the luminosity. Furthermore, the statistical uncertainty scales as $1/\sqrt{\sigmaopt(\thetaj)}$ and thus depends on the event rate in the \thetaj bin in which the energy asymmetry is measured.

To obtain numerical values for the uncertainties, we adopt the selection efficiencies and background estimates from a recent measurement of differential cross sections in inclusive \ttbar production at the ATLAS experiment~\cite{CERN-EP-2019-149}, which has similar particle-level selection criteria to ours. Assuming constant efficiencies and background uncertainties
\begin{align}
\effReco(\thetaj)=45\%,\quad \effPart(\thetaj)=80\%,\quad \fttbar(\thetaj)=85\%,\quad \Delta \Nbkg/\Nbkg=10\%,
\end{align}
we obtain compact estimates of the statistical and background observables
\begin{equation}\label{eq:exp-unc}
\Delta A_E^{\text{stat}}(\thetaj)\approx\frac{1.4}{ \sqrt{L\,\sigma_{t\bar t j}^{\rm opt}(\thetaj)}}\,,\qquad 
\Delta A_E^{\text{bkg}}(\thetaj)\approx 0.018\cdot|A_{E,\mathrm{meas}}^{\text{opt}}(\thetaj)|\,.
\end{equation}

\section{Bounds on individual Wilson coefficients}\label{tables_bounds}
\noindent In this appendix, we give numerical results for the expected bounds on individual Wilson coefficients from one-parameter fits to the energy asymmetry and the cross section in $t\bar t j$ production at 68\% CL (Table~\ref{tab:68bounds}) and 95\% CL (Table~\ref{tab:95bounds}). We also show existing bounds from a measurement of the rapidity asymmetry $A_{|y|}$ in $t\bar t$ production, as well as various fits to combinations of these observables. The results are visualized in Figure~\ref{fig:indiv_combi}.

\begin{table}[h]
	\small
	\begin{tabular}{c|cccc}
		\noalign{\hrule height 1pt}
	coefficient	& $\phantom{\Big[}$  $A_{E}^{1-3}$ & $A_{E}^{1}$ & $A_{E}^{2}$ & $A_{E}^{3}$ \\
		\hline
$\phantom{\Big[}$ $C_{tG}\ $ & [-1.5,3.6] & [ -- , -- ] & [-1.6,3.5] & [ -- , -- ] \\
$\phantom{\Big[}$ $C_{Qq}^{3,8}\ $ & [-0.9,0.8] & [-1.9,1.5] & [-1.0,1.4] & [-1.1,0.9]\\
$\phantom{\Big[}$ $C_{Qq}^{1,8}\ $ & [-0.8,0.7] & [-2.3,1.2] & [-0.7,1.9] & [-1.5,0.7] \\
$\phantom{\Big[}$ $C_{Qu}^{8}\ $ & [-1.4,0.9] & [-1.8,1.6] & [-1.8,7.0] & [-1.8,0.9] \\
$\phantom{\Big[}$ $C_{Qd}^{8}\ $ & [-2.1,1.5] & [-2.6,2.3] & [-3.9,15.3] & [-2.4,1.5] \\
$\phantom{\Big[}$ $C_{tq}^{8}\ $ & [-1.2,0.7] & [-1.5,1.3] & [-1.3,7.9] & [-1.6,0.7] \\
$\phantom{\Big[}$ $C_{tu}^{8}\ $ & [-1.0,0.9] & [-2.7,1.5] & [-0.9,2.1] & [-1.6,0.9] \\
$\phantom{\Big[}$ $C_{td}^{8}\ $ & [-1.7,1.4] & [-3.5,2.5] & [-1.7,2.8] & [-2.2,1.5] \\
$\phantom{\Big[}$ $C_{Qq}^{3,1}\ $ & [-0.4,0.4] & [-0.6,0.6] & [-0.6,0.7] & [-0.4,0.4] \\
$\phantom{\Big[}$ $C_{Qq}^{1,1}\ $ & [-0.4,0.4] & [-0.7,0.6] & [-0.5,0.7] & [-0.4,0.4] \\
$\phantom{\Big[}$ $C_{Qu}^{1}\ $ & [-0.5,0.5] & [-0.6,0.8] & [-1.6,1.1] & [-0.5,0.6] \\
$\phantom{\Big[}$ $C_{Qd}^{1}\ $ & [-0.7,0.8] & [-1.0,1.1] & [-3.0,2.3] & [-0.8,0.8] \\
$\phantom{\Big[}$ $C_{tq}^{1}\ $ & [-0.4,0.4] & [-0.5,0.7] & [-1.4,1.0] & [-0.4,0.5] \\
$\phantom{\Big[}$ $C_{tu}^{1}\ $ & [-0.5,0.5] & [-0.8,0.7] & [-0.6,0.9] & [-0.5,0.5] \\
$\phantom{\Big[}$ $C_{tu}^{1}\ $ & [-0.7,0.8] & [-1.2,1.1] & [-1.0,1.3] & [-0.8,0.8] \\
		\noalign{\hrule height 1pt}
	\end{tabular}\\[0.2cm]
	
	\begin{tabular}{c|ccccc}
		\noalign{\hrule height 1pt}
	coefficient	& $\phantom{\Big[}$  $\sigma_{t\bar{t}j}$ & $A_{{E}}^{{1-3}} \& \sigma_{t\bar{t}j}$ & $A_{|y|}$ & $A_{{E}}^{{1-3}} \& A_{|y|}$ & $A_{{E}}^{{1-3}} \& A_{|y|} \& \sigma_{t\bar{t}j}$ \\
		\hline
$\phantom{\Big[}$ $C_{tG}\ $ & [-2.8,0.7] & [-1.3,0.7] & [ -- ,3.4] & [-1.2,2.4] & [-1.0,0.7]\\
$\phantom{\Big[}$ $C_{Qq}^{3,8}\ $ & [-2.8,2.7] & [-0.9,0.8] & [-2.4,1.3] & [-0.9,0.8] & [-0.9,0.8]\\
$\phantom{\Big[}$ $C_{Qq}^{1,8}\ $ & [-3.1,2.5] & [-0.8,0.7] & [-3.6,0.8] & [-0.7,0.6] & [-0.7,0.6]\\
$\phantom{\Big[}$ $C_{Qu}^{8}\ $ & [-3.4,2.8] & [-1.4,0.9] &  [-4.6,1.5] & [-1.4,0.9] & [-1.4,0.9]\\
$\phantom{\Big[}$ $C_{Qd}^{8}\ $ & [-4.8,4.1] & [-2.1,1.5] & [-5.2,3.2] & [-2.1,1.5] & [-2.1,1.5]\\
$\phantom{\Big[}$ $C_{tq}^{8}\ $ & [-2.8,2.2] & [-1.1,0.7] & [-3.7,1.3] & [-1.1,0.7] & [-1.1,0.7]\\
$\phantom{\Big[}$ $C_{tu}^{8}\ $ & [-3.6,3.1] & [-1.0,0.9] & [-3.4,1.2] & [-1.0,0.8] & [-1.0,0.8]\\
$\phantom{\Big[}$ $C_{td}^{8}\ $ & [-5.2,4.6] & [-1.7,1.4] & [-5.1,1.9] & [-1.7,1.4] & [-1.6,1.4]\\
$\phantom{\Big[}$ $C_{Qq}^{3,1}\ $ & [-1.1,1.1] & [-0.4,0.4] & [-0.6,0.7] & [-0.4,0.4] & [-0.4,0.4]\\
$\phantom{\Big[}$ $C_{Qq}^{1,1}\ $ & [-1.0,1.1] & [-0.4,0.4] & [-0.7,0.6] & [-0.4,0.4] & [-0.4,0.4]\\
$\phantom{\Big[}$ $C_{Qu}^{1}\ $ & [-1.3,1.3] & [-0.5,0.5] & [-0.9,1.0] & [-0.5,0.5] & [-0.4,0.5]\\
$\phantom{\Big[}$ $C_{Qd}^{1}\ $ & [-1.9,1.9] & [-0.7,0.8] & [-1.4,1.5] & [-0.7,0.8] & [-0.7,0.8]\\
$\phantom{\Big[}$ $C_{tq}^{1}\ $ & [-1.1,1.1] & [-0.4,0.4] & [-0.7,0.9] & [-0.4,0.4] & [-0.4,0.4]\\
$\phantom{\Big[}$ $C_{tu}^{1}\ $ & [-1.3,1.4] & [-0.5,0.5] & [-0.9,0.7] & [-0.5,0.5] & [-0.4,0.5]\\
$\phantom{\Big[}$ $C_{tu}^{1}\ $ & [-1.9,2.0] & [-0.7,0.7] & [-1.2,1.3] & [-0.7,0.7] & [-0.7,0.7]\\
		\noalign{\hrule height 1pt}
	\end{tabular}
	\caption{Expected 68\% CL bounds on individual Wilson coefficients from fits to LHC measurements of the optimized energy asymmetry in three $\theta_j$ bins $A_E^1$, $A_E^2$, $A_E^3$, their combination $A_E^{1-3}$, and the cross section $\sigma_{t\bar t j}$. All $t\bar t j$ observables are based on the boosted phase-space selection and a data set of $139\,\text{fb}^{-1}$. We also show existing bounds from the rapidity asymmetry $A_{|y|}$ in $t\bar t$ production measured during Run 2~\cite{ATLAS-CONF-2019-026}, as well as combined fits to several observables. Empty spaces indicate that no limits could be found within $C \in [-25,25]$. Bounds on $C_i$ are reported in units of TeV$^{-2}$.\label{tab:68bounds}}
\end{table}

\begin{table}[h]
	\small
	\begin{tabular}{c|cccc}
			\noalign{\hrule height 1pt}
		 coefficient	&  $\phantom{\Big[}$ $A_{E}^{1-3}$ & $A_{E}^{1}$ & $A_{E}^{2}$ & $A_{E}^{3}$ \\
\hline
$\phantom{\Big[}$ $C_{tG}\ $ & [-2.6,22.3] & [ -- , -- ] & [-2.8,20.6] & [ -- , -- ] \\
$\phantom{\Big[}$ $C_{Qq}^{3,8}\ $ & [-1.3,1.3] & [-2.7,2.3] & [-1.5,2.0] & [-1.6,1.3] \\
$\phantom{\Big[}$ $C_{Qq}^{1,8}\ $ & [-1.2,1.1] & [-3.2,2.0] & [-1.2,2.5] & [-1.9,1.1] \\
$\phantom{\Big[}$ $C_{Qu}^{8}\ $ & [-2.0,1.5] & [-2.6,2.4] & [-3.5,14.6] & [-2.4,1.5] \\
$\phantom{\Big[}$ $C_{Qd}^{8}\ $ & [-2.9,2.4] & [-3.7,3.5] & [-11.7, -- ] & [-3.3,2.4] \\
$\phantom{\Big[}$ $C_{tq}^{8}\ $ & [-1.7,1.2] & [-2.1,1.9] & [-2.7,22.9] & [-2.0,1.2] \\
$\phantom{\Big[}$ $C_{tu}^{8}\ $ & [-1.5,1.4] & [-3.7,2.5] & [-1.5,2.7] & [-2.1,1.4] \\
$\phantom{\Big[}$ $C_{td}^{8}\ $ & [-2.4,2.2] & [-4.9,3.8] & [-2.7,3.9] & [-3.0,2.2] \\
$\phantom{\Big[}$ $C_{Qq}^{3,1}\ $ & [-0.6,0.6] & [-0.9,0.9] & [-0.9,1.0] & [-0.6,0.6] \\
$\phantom{\Big[}$ $C_{Qq}^{1,1}\ $ & [-0.6,0.6] & [-1.0,0.9] & [-0.8,1.1] & [-0.6,0.6] \\
$\phantom{\Big[}$ $C_{Qu}^{1}\ $ & [-0.7,0.8] & [-1.0,1.2] & [-2.8,2.1] & [-0.7,0.8] \\
$\phantom{\Big[}$ $C_{Qd}^{1}\ $ & [-1.1,1.1] & [-1.4,1.6] & [-8.6,6.1] & [-1.1,1.2] \\
$\phantom{\Big[}$ $C_{tq}^{1}\ $ & [-0.6,0.6] & [-0.8,1.0] & [-2.8,1.9] & [-0.6,0.7] \\
$\phantom{\Big[}$ $C_{tu}^{1}\ $ & [-0.7,0.7] & [-1.2,1.1] & [-1.0,1.3] & [-0.7,0.8] \\
$\phantom{\Big[}$ $C_{tu}^{1}\ $ & [-1.0,1.1] & [-1.7,1.6] & [-1.6,1.9] & [-1.1,1.2] \\
		\noalign{\hrule height 1pt}
	\end{tabular}\\[0.2cm]
	\begin{tabular}{c|ccccc}
			\noalign{\hrule height 1pt}
		 coefficient	&  $\phantom{\Big[}$ $\sigma_{t\bar{t}j}$ & $A_{{E}}^{{1-3}} \& \sigma_{t\bar{t}j}$ & $A_{|y|}$ & $A_{{E}}^{{1-3}} \& A_{|y|}$ & $A_{{E}}^{{1-3}} \& A_{|y|} \& \sigma_{t\bar{t}j}$ \\
\hline
$\phantom{\Big[}$ $C_{tG}\ $ & [-3.3,1.2] & [-2.5,1.2] & [ -- , -- ] & [-2.1,10.6] & [-2.1,1.1]\\
$\phantom{\Big[}$ $C_{Qq}^{3,8}\ $ & [-4.0,3.9] & [-1.3,1.3] & [-3.4,2.2] & [-1.3,1.2] & [-1.3,1.2]\\
$\phantom{\Big[}$ $C_{Qq}^{1,8}\ $ & [-4.2,3.7] & [-1.2,1.1] & [-4.3,1.5] & [-1.2,1.0] & [-1.2,1.0]\\
$\phantom{\Big[}$ $C_{Qu}^{8}\ $ & [-4.6,4.1] & [-2.0,1.5] & [-5.5,2.4] & [-2.0,1.4] & [-2.0,1.4]\\
$\phantom{\Big[}$ $C_{Qd}^{8}\ $ & [-6.7,6.0] & [-2.9,2.3] & [-6.7,4.7] & [-2.9,2.3] & [-2.9,2.3]\\
$\phantom{\Big[}$ $C_{tq}^{8}\ $ & [-3.9,3.3] & [-1.7,1.2] & [-4.5,2.1] & [-1.7,1.1] & [-1.7,1.1]\\
$\phantom{\Big[}$ $C_{tu}^{8}\ $ & [-5.0,4.5] & [-1.5,1.4] & [-4.5,2.2] & [-1.5,1.3] & [-1.5,1.3]\\
$\phantom{\Big[}$ $C_{td}^{8}\ $ & [-7.2,6.6] & [-2.4,2.2] & [-6.6,3.3] & [-2.4,2.1] & [-2.4,2.1]\\
$\phantom{\Big[}$ $C_{Qq}^{3,1}\ $ & [-1.5,1.6] & [-0.6,0.6] & [-1.0,1.1] & [-0.6,0.6] & [-0.5,0.6]\\
$\phantom{\Big[}$ $C_{Qq}^{1,1}\ $ & [-1.5,1.6] & [-0.6,0.6] & [-1.0,1.0] & [-0.5,0.6] & [-0.5,0.6]\\
$\phantom{\Big[}$ $C_{Qu}^{1}\ $ & [-1.9,1.9] & [-0.7,0.8] & [-1.3,1.4] & [-0.7,0.8] & [-0.7,0.8]\\
$\phantom{\Big[}$ $C_{Qd}^{1}\ $ & [-2.8,2.8] & [-1.1,1.1] & [-2.0,2.0] & [-1.0,1.1] & [-1.0,1.1]\\
$\phantom{\Big[}$ $C_{tq}^{1}\ $ & [-1.5,1.6] & [-0.6,0.6] & [-1.0,1.2] & [-0.6,0.6] & [-0.5,0.6]\\
$\phantom{\Big[}$ $C_{tu}^{1}\ $ & [-1.9,1.9] & [-0.7,0.7] & [-1.4,1.2] & [-0.7,0.7] & [-0.7,0.7]\\
$\phantom{\Big[}$ $C_{tu}^{1}\ $ & [-2.7,2.8] & [-1.0,1.1] & [-1.9,2.0] & [-1.0,1.1] & [-1.0,1.1]\\
		\noalign{\hrule height 1pt}
	\end{tabular}
	\caption{Expected 95\% CL bounds on individual Wilson coefficients from fits to LHC measurements of the optimized energy asymmetry in three $\theta_j$ bins $A_E^1$, $A_E^2$, $A_E^3$, their combination $A_E^{1-3}$, and the cross section $\sigma_{t\bar t j}$. All $t\bar t j$ observables are based on the boosted phase-space selection and a data set of $139\,\text{fb}^{-1}$. We also show existing bounds from the rapidity asymmetry $A_{|y|}$ in $t\bar t$ production measured during Run 2~\cite{ATLAS-CONF-2019-026}, as well as combined fits to several observables.	Empty spaces indicate that no limits could be found within $C \in [-25,25]$.  Bounds on $C_i$ are reported in units of TeV$^{-2}$.\label{tab:95bounds}}
\end{table}

\bibliographystyle{JHEP}

\end{document}